\pdfoutput=1

\documentclass[11pt,twoside,a4paper,cmspaper,final,collab]{cms-tdr}
\def\svnVersion{298338}\def\cmsCernNo{2013-037}\def\cmsCernDate{\today}\def\cmsMessage{Published in the European Physical Journal C as \href{http://dx.doi.org/10.1140/epjc/s10052-015-3533-3}{\doi{10.1140/epjc/s10052-015-3533-3.}}}

\begin{document}\cmsNoteHeader{EXO-13-006}

\hyphenation{had-ron-i-za-tion}
\hyphenation{cal-or-i-me-ter}
\hyphenation{de-vices}
\RCS$Revision: 290109 $
\RCS$HeadURL: svn+ssh://alverson@svn.cern.ch/reps/tdr2/papers/EXO-13-006/trunk/EXO-13-006.tex $
\RCS$Id: EXO-13-006.tex 290109 2015-05-26 00:44:02Z alverson $
\newlength\cmsFigWidth
\ifthenelse{\boolean{cms@external}}{\setlength\cmsFigWidth{0.85\columnwidth}}{\setlength\cmsFigWidth{0.4\textwidth}}
\ifthenelse{\boolean{cms@external}}{\providecommand{\cmsLeft}{top}}{\providecommand{\cmsLeft}{left}}
\ifthenelse{\boolean{cms@external}}{\providecommand{\cmsRight}{bottom}}{\providecommand{\cmsRight}{right}}
\providecommand{\dEdx}{\ensuremath{\rd{E}\hspace{-2pt}/\hspace{-2pt}\rd{x}}\xspace}
\newcommand{\mthresh}{\ensuremath{m_\text{thresh}}\xspace}

\cmsNoteHeader{EXO-13-006}
\title{Constraints on the pMSSM, AMSB model and on other models from the search for long-lived charged particles in proton-proton collisions at $\sqrt{s} =8\TeV$}
\titlerunning{Constraints on beyond standard models from a search for long-lived charged particles}

\date{\today}

\abstract{
Stringent limits are set on the long-lived lepton-like sector of the phenomenological minimal
supersymmetric standard model (pMSSM) and the anomaly-mediated supersymmetry breaking (AMSB) model.
The limits are derived from the results presented in a recent search for long-lived charged particles in proton-proton collisions, based on data collected by the CMS detector at a centre-of-mass energy of 8\TeV at the Large Hadron Collider.
In the pMSSM parameter sub-space considered, 95.9\% of the points predicting charginos with a lifetime of at least 10\unit{ns} are excluded.
These constraints on the pMSSM are the first obtained at the LHC.
Charginos with a lifetime greater than 100\unit{ns} and masses up to about 800\GeV in the AMSB model are also excluded.
The method described can also be used to set constraints on other models.
}

\hypersetup{%
pdfauthor={CMS Collaboration},%
pdftitle={Constraints on the pMSSM, AMSB model and on other models from the search for long-lived charged particles in proton-proton collisions at sqrt(s) =  8 TeV},%
pdfsubject={CMS},%
pdfkeywords={CMS, physics, BSM, long-lived particles}}

\maketitle

\section{Introduction}

{\tolerance=400
We present new constraints on long-lived chargino production in the phenomenological minimal supersymmetric standard model (pMSSM)~\cite{Djouadi:1998di} and on the anomaly-mediated supersymmetry breaking (AMSB) model \cite{Chen:1996ap, Giudice1998, Randall1999}.
In the pMSSM, we consider the parameter sub-space for particle masses up to 3\TeV and charginos (\PSGcpm) with a mean proper decay length ($c\tau)$ greater than 50\cm.
In the AMSB model, the small mass difference between the lightest chargino and neutralino (\PSGczDo) often leads to a long chargino lifetime.
\par}

Long-lived charged particles are predicted by various extensions of the standard model (SM)~\cite{StandardModel67_1,StandardModel67_2,StandardModel67_3}, such as supersymmetry (SUSY)~\cite{Nilles19841} and theories with
extra dimensions~\cite{ArkaniHamed1998263,PhysRevLett.83.3370}.
If such particles have a mass lighter than a few \TeV they could be produced by the CERN LHC.
The energy available in the proton-proton (pp) collisions at the LHC is such that particles with mass $\gtrsim$100\GeV and lifetime greater than $\mathcal{O}(1)$\unit{ns} could be observed with the CMS detector~\cite{Chatrchyan:2008zzk} as high-momentum tracks with anomalously large rates of energy loss
through ionization (\dEdx). These particles could also be highly penetrating such that the fraction reaching the CMS muon system would be sizable.
The muon system could therefore be used to help in identification and in the measurement of the time-of-flight (TOF) of the particles.  	
The signature described above is exploited in a previous CMS search~\cite{EXO-12-026}, which sets the most stringent limits to date on a number of representative models predicting
massive long-lived charged particles such as tau sleptons (staus), top squarks, gluinos, and leptons with an electric charge between $e/3$ and $8e$.

The main thrust of this paper is to present constraints on the pMSSM and the AMSB model, obtained using the results from the search for heavy stable charged particles (HSCP) in proton-proton collisions at  $\sqrt{s} = 7$ and 8\TeV~\cite{EXO-12-026}.
The method applied relies on the factorization of the acceptance in terms of two probabilities, calculated using the standard CMS simulation and reconstruction tools at 8\TeV.
Moreover, the results of the acceptance calculations have been tabulated and made publicly available~\cite{ProbMaps}.
Thus this technique, which allows the signal acceptance for a model with long-lived particles to be computed using the kinematic properties of the particles at their production point, may in the future be used by others to evaluate constraints on other extensions of the SM without use of the CMS software.

\section{The CMS detector}

The central feature of the CMS apparatus
is a superconducting solenoid of 6\unit{m} internal diameter providing a field of 3.8\unit{T}.
Within the field volume are
a silicon pixel and strip tracker, a lead tungstate crystal
electromagnetic calorimeter, and a brass and scintillator
hadron calorimeter. Muons are measured in gas-ionization
detectors embedded in the steel flux-return yoke of the magnet.
The tracker measures charged particles within the pseudorapidity range $\abs{\eta}< 2.5$ and provides a transverse momentum \pt resolution of about 2.8\% for 100\GeV particles.
The analog readout of the tracker also enables the particle ionization energy loss to be measured with a resolution of about 5\%.
Muons are measured in the range $\abs{\eta}< 2.4$, with
detection planes based on three technologies: drift tubes,
cathode strip chambers, and resistive-plate chambers.
The muon system extends out to 11\unit{m} from the interaction point in the $z$ direction and
to 7\unit{m} radially.
Matching tracks in the muon system to tracks measured in the silicon tracker results in a transverse
momentum resolution between 1 and 5\%, for \pt values up to 1\TeV.
The time resolution of the muon system is of the order of 1\unit{ns}.
This provides a time-of-flight measurement that can be used to determine the inverse of the long-lived particle velocity as a fraction of the velocity of light ($1/\beta$) with a resolution of 0.065 over the full $\eta$ range~\cite{EXO-12-026}.
A more detailed description of the CMS detector, of the coordinate system, and of the kinematic variables can be found in Ref.~\cite{Chatrchyan:2008zzk}.

The first level of the CMS trigger system, composed of custom
hardware processors, uses information from the calorimeters and
muon detectors to select the events of interest.
The high-level trigger processor
farm further decreases the event rate from around 100\unit{kHz} to
around 400\unit{Hz} for data storage.

\section{Estimation of signal acceptance}
\label{sec:techniqueIntro}

Transcribing the results presented in Ref.~\cite{EXO-12-026},
in terms of limits on models other than those considered in the reference,
requires the tabulation of the relevant signal acceptance.
This acceptance can then be used to constrain the beyond the standard model (BSM) scenarios, as described in the next sections.
There can be significant differences in signal acceptance between the models investigated
in Ref.~\cite{EXO-12-026} and the model to be tested. These differences arise because
the \dEdx and TOF measurements are affected by the distribution and orientation of the material encountered by particles travelling within the CMS detector.
The combination of these effects with the differences in the kinematic properties between
models can result in large differences in signal acceptance.

The signal acceptance can be accurately computed by
fully  simulating  and reconstructing signal events in the CMS
detector and by applying the same selection criteria as adopted in Ref.~\cite{EXO-12-026}.
We refer to this method of determining the signal acceptance as the ``full simulation''.
This procedure requires extensive knowledge of the CMS detector and in particular use of CMS software that accurately models detector response to simulated signal events and employs the full CMS reconstruction routines such that identical selection criteria can be used on simulated signal events as on data collected from collisions.

An alternate method, which only requires information on the kinematic properties
of the long-lived particles at their production point, is presented here. Such a method can be used if the
efficiency for triggering and selecting events can be expressed in terms of probabilities
associated with each individual long-lived particle.
This is the case for models
with lepton-like massive long-lived particles since the event selection specified in Ref.~\cite{EXO-12-026} imposes only requirements on measurements performed on
individual particles.  The adjective ``lepton-like", defined in Ref.~\cite{EXO-12-026}, indicates particles that do not interact strongly and are therefore not subject to hadronization.

The event selection requirements of Ref.~\cite{EXO-12-026} are expressed in terms of measured
\pt, \dEdx, TOF, and mass values of individual particles in an event.
The probability that a long-lived particle in an event passes the online or offline selection requirements in
Ref.~\cite{EXO-12-026} can be expressed as a function of the true (generator-level) kinematic properties ($k$) of the particle: $\beta$, $\eta$, and \pt.
Any other equivalent set of kinematic and directional variables could also be used to express this dependence.
The offline selection of Ref.~\cite{EXO-12-026} has fixed values for \pt, \dEdx, and TOF thresholds but uses different
reconstructed mass thresholds (\mthresh) depending on the mass ($m$) of the long-lived particle in the model being tested.

With these individual particle probabilities, the acceptance
$\mathcal{A}$ for a model to pass the online and offline
selections can be computed with a Monte Carlo technique by
generating a large number of events $N$ such that
\begin{equation}
\label{eq:Technique}
\mathcal{A} = \frac{1}{N} \sum_{i}^{N} P^{\text{on}}(k^1_i, \dots, k^M_i) \times
P^{\text{off}}(\mthresh, k^1_i, \dots, k^M_i),
\end{equation}
where $P^{\text{on}}(k^1_i, k^2_i, \dots, k^M_i)$ is the probability that the event
with index $i$ containing $M$ long-lived particles with true kinematic properties
$k^1_i, k^2_i, \dots, k^M_i$ passes the online selection, and
$P^{\text{off}}(\mthresh,k^1_i, k^2_i, \dots, k^M_i)$  is the probability that
the event with the same kinematic properties passes the offline selection
with mass threshold \mthresh, after having passed the online selection.
Given the mass resolution of the detector, a mass threshold of $\mthresh \simeq 0.6 m$ has an efficiency of about 95\% for the benchmark models considered in Ref.~\cite{EXO-12-026}.
Consequently, throughout this paper, mass thresholds of 0, 100, 200, and 300\GeV are used for true long-lived particle masses of  $m \leq 166$, $\leq$330, $\leq$500, and $\ge$500\GeV, respectively.
The choice of 100\GeV steps for the mass thresholds is made so that the information provided in Table 3 of Ref.~\cite{EXO-12-026} regarding the background expectation and the observed count in the signal region may be used, thus allowing a more general application of the factorization method.
In the case where only one long-lived particle is present in each
event, the probabilities have the simplest form $P^{\text{on}}(k)$ and
$P^{\text{off}}(\mthresh, k)$.
If each  event contains two long-lived particles, the probabilities can be expressed using the probabilities for events
with either single long-lived particle:
\ifthenelse{\boolean{cms@external}}{
\begin{multline}
\label{eq:EventAcceptance}
\shoveleft{P^{\text{on}}(k^1_i, k^2_i) = P^{\text{on}}(k^1_i)  + P^{\text{on}}(k^2_i) - P^{\text{on}}(k^1_i)  P^{\text{on}}(k^2_i)  ;} \\
\shoveleft{P^{\text{off}}(\mthresh, k^1_i, k^2_i)  = P^{\text{off}}(\mthresh, k^1_i)  +}\\
\shoveright{P^{\text{off}}(\mthresh, k^2_i) - P^{\text{off}}(\mthresh, k^1_i)
P^{\text{off}}(\mthresh, k^2_i) .}
\end{multline}
}{
\begin{equation}
\label{eq:EventAcceptance}
\begin{split}
P^{\text{on}}(k^1_i, k^2_i) &= P^{\text{on}}(k^1_i)  + P^{\text{on}}(k^2_i) - P^{\text{on}}(k^1_i)   P^{\text{on}}(k^2_i)  ; \\
P^{\text{off}}(\mthresh, k^1_i, k^2_i) & = P^{\text{off}}(\mthresh, k^1_i)  +
P^{\text{off}}(\mthresh, k^2_i) - P^{\text{off}}(\mthresh, k^1_i)
P^{\text{off}}(\mthresh, k^2_i) .
\end{split}
\end{equation}
}
The expression for events with more than two long-lived particles per event can also be expressed in terms of the probabilities $P^{\text{on}}(k)$ and
$P^{\text{off}}(\mthresh, k)$ associated with each individual long-lived particle.

The probabilities $P^{\text{on}}(k)$ and $P^{\text{off}}(\mthresh, k)$, derived using the full
simulation as described in the Appendix, are evaluated using the
selection requirements adopted by the ``tracker+TOF''
analysis described in Ref.~\cite{EXO-12-026}.
This selection, where tracks are required to be reconstructed in both
the tracker and the muon system, has been shown to be the most sensitive
to signatures with lepton-like long-lived particles.
The probabilities thus obtained are stored in the form of look-up tables.

The factorization method described above is validated by comparing the estimated signal
acceptance values with those obtained with full simulation for a few benchmark models predicting long-lived leptons.
In the rest of this paper we refer to the former as ``the fast technique".
In the full simulation case, pileup due to multiple interactions per bunch crossing is also simulated.
Agreement better than 10\% between full simulation and the fast technique presented in this paper is generally observed for the considered values of \mthresh.

More details on the determination of the probabilities $P^{\text{on}}(k)$ and $P^{\text{off}}(\mthresh, k)$ can be found in the Appendix, which also contains the details of the validation of the fast technique based on Eq.~(\ref{eq:Technique}).
The Appendix also explains how this technique can be used in the future to estimate the CMS exclusion limits for extensions of the SM other than the ones considered in this paper using publicly available look-up tables for $P^{\text{on}}(k)$ and $P^{\text{off}}(\mthresh, k)$.

\begin{figure}[tbhp]
 \begin{center}
  \includegraphics[width=0.49\textwidth]{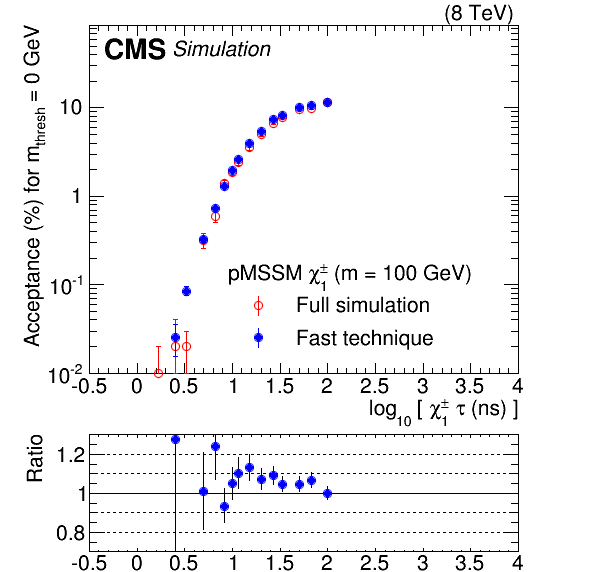}
  \includegraphics[width=0.49\textwidth]{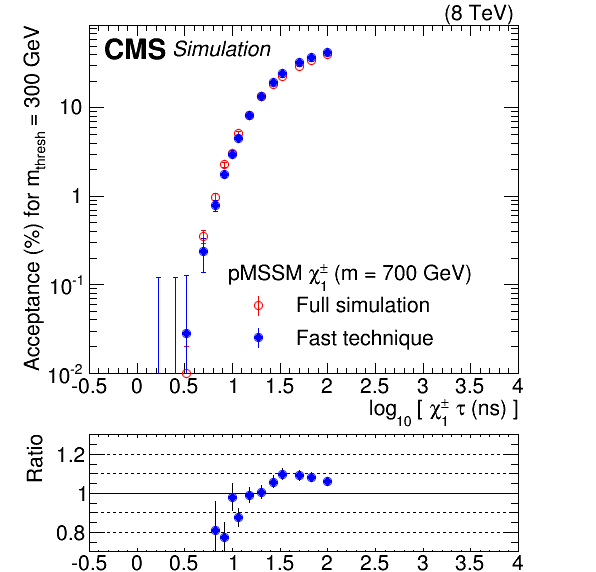}
 \end{center}
 \caption{Signal acceptance as a function of the chargino lifetime for a benchmark model having a chargino of mass 100\GeV (\cmsLeft) and 700\GeV (\cmsRight), with a mass threshold of 0\GeV and 300\GeV, respectively.
    The panel below each figure shows the ratio of acceptance from the fast technique to the acceptance obtained from a full simulation of the detector.
    \label{fig:TkTOFpMSSMValidationUnstable}}
\end{figure}

The probabilities $P^{\text{on}}(k)$ and $P^{\text{off}}(\mthresh, k)$ are
computed with simulated stable particles, but the method is easily
extended to particles with finite lifetimes by correcting the $P^{\text{on}}(k)$ probability for the probability that the
long-lived particle, with mass $m$, lifetime $\tau$, and momentum $p$, travels at least
the distance $x$ required to produce the minimum number of track
measurements in the CMS muon system, as required in Ref.~\cite{EXO-12-026}.
The correction consists of an exponential factor, $\exp[ -m x/(\tau p)]$, to be applied to $P^{\text{on}}(k)$.
This correction is not applied also to  $P^{\text{off}}(\mthresh, k)$, since it needs to be made just once for each long-lived particle.
The distance $x$ only depends on the pseudorapidity of the particle:
\begin{equation}
x =
\begin{cases}
 9.0\unit{m} & 0.0 \leq \abs{\eta} \leq 0.8; \\
10.0\unit{m} & 0.8 \leq \abs{\eta} \leq 1.1; \\
11.0\unit{m} & 1.1 \leq \abs{\eta}.
\end{cases}
\end{equation}
These values of $x$ ensure that the particle traverses the entire muon system before decaying.
This choice results in a conservative estimate of the signal acceptance since it ignores the contribution to the acceptance from particles
that decay before the end of the muon detector but still pass the selection.
In Fig.~\ref{fig:TkTOFpMSSMValidationUnstable}, the acceptance obtained with the fast technique is compared with the acceptance obtained with the full simulation of the detector, as a function of the lifetime of the particle.
It can be seen from the ratio panels that in most cases the agreement between the two methods is within 10\%, corresponding to the systematic uncertainty in the fast estimation technique used to compute the acceptance.
 For lifetimes less than 10\unit{ns}, the spread is somewhat larger, but the tendency to underestimate the acceptance is less than 15\%.

\begin{figure*}[htb]
 \begin{center}
  \includegraphics[width=0.49\textwidth]{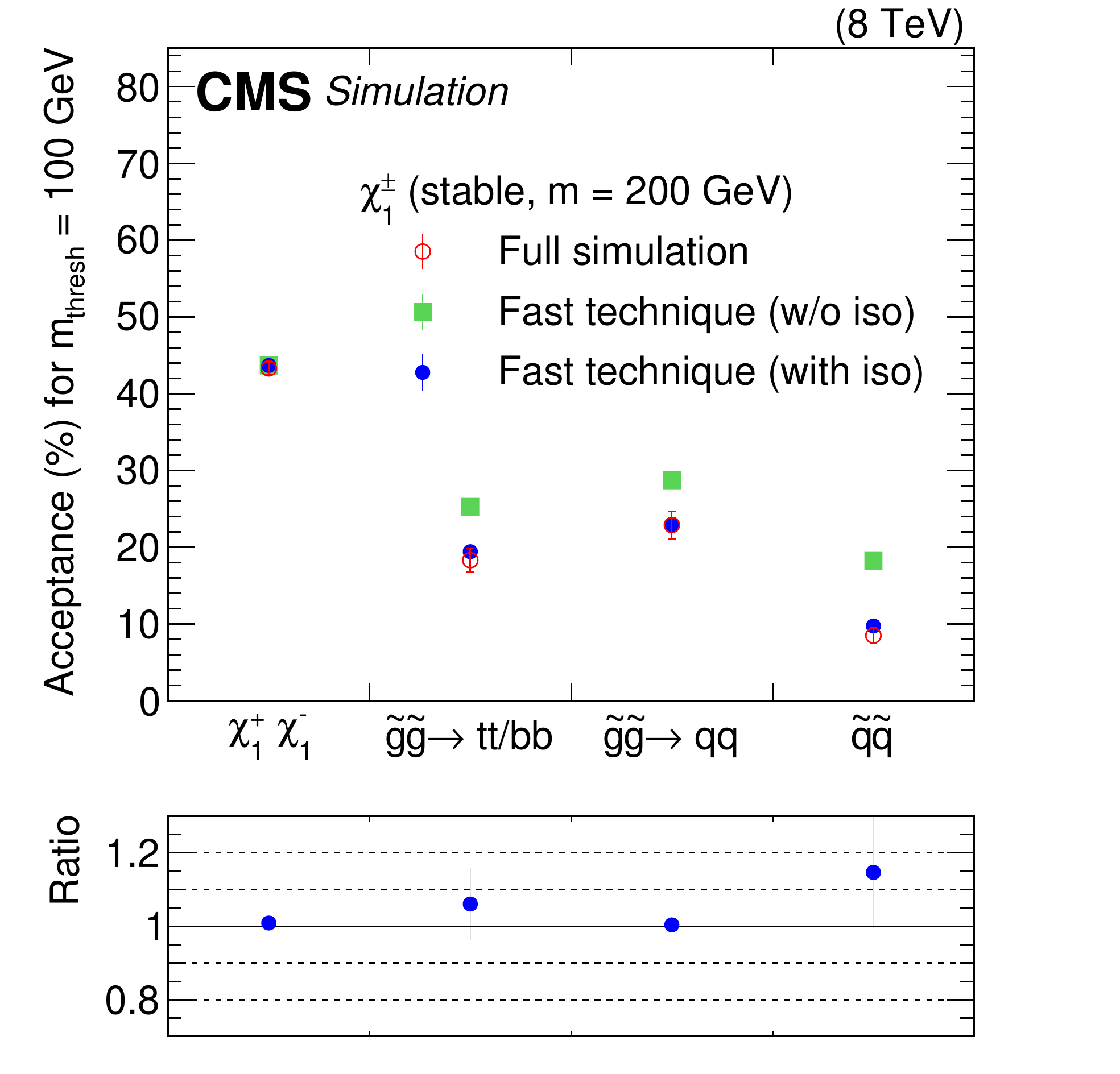}
  \includegraphics[width=0.49\textwidth]{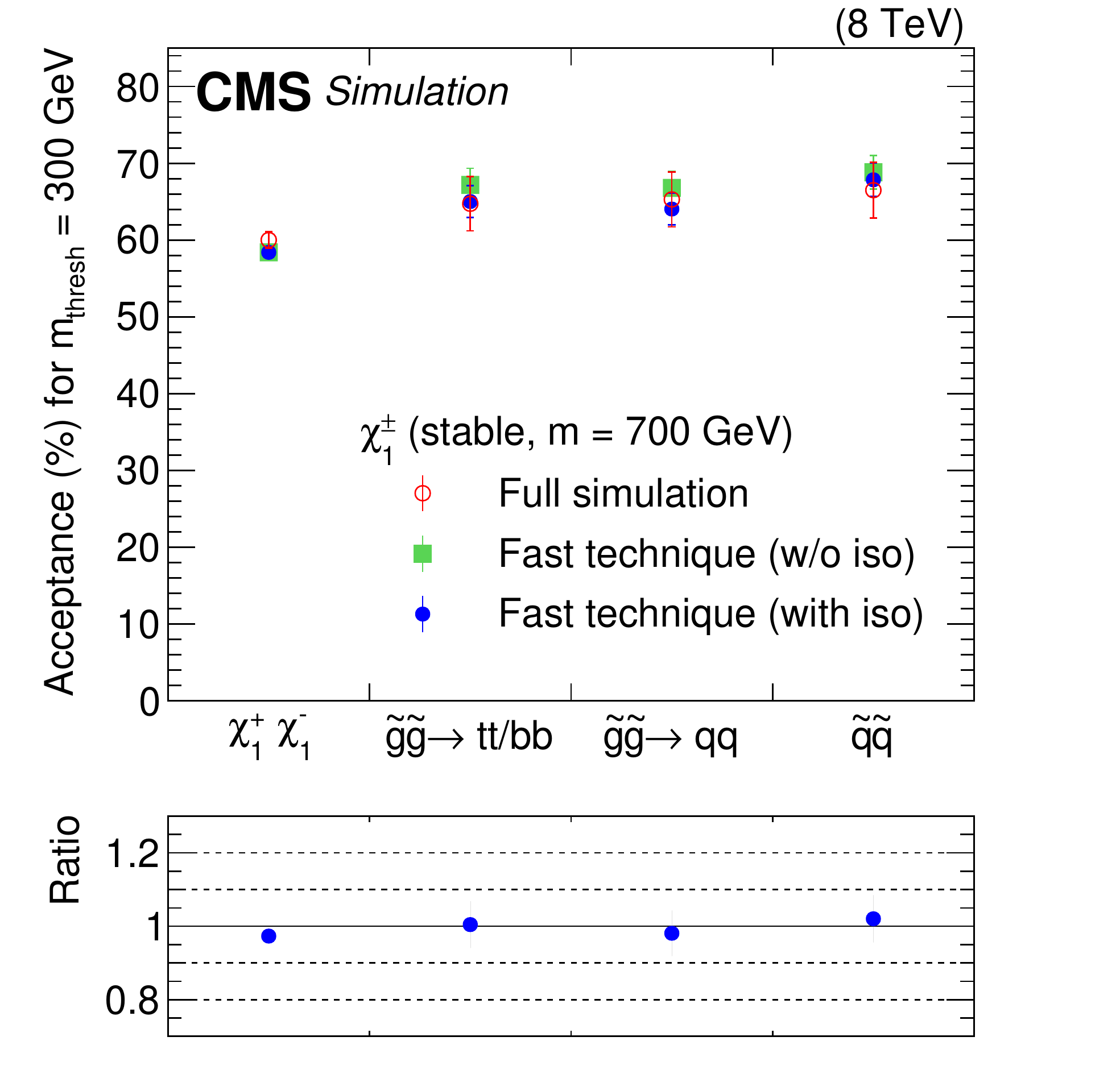}
 \end{center}
 \caption{Signal acceptance as a function of the chargino production mechanism for a benchmark model having a chargino of mass 200\GeV (left) and 700\GeV (right), with a mass threshold of 100\GeV and 300\GeV, respectively.
    From left to right, the production mechanisms considered are: direct pair production of charginos; pair production of gluinos that each decay to a heavy quark (b,t) and a chargino; pair production of gluinos that each decay to a light quark (u,d,s,c) and a chargino; and pair production of squarks that each decay to a quark and a chargino.
    The panel below each figure shows the ratio of acceptance from the fast technique with the isolation requirements to the acceptance obtained from a full simulation of the detector.
    The estimated acceptance is given with and without the generator-level isolation.  Pileup is present only in the full simulation samples.
    \label{fig:TkTOFpMSSMValidationIsolation}}
\end{figure*}

The offline event selection in Ref.~\cite{EXO-12-026} includes two
isolation requirements. The first is defined  by $\Sigma \pt <
50$\GeV, where the sum is over all tracks (except the candidate
track) within a radius $\Delta R =
\sqrt{\smash[b]{(\Delta \eta)^2 + (\Delta \phi)^2}} = 0.3$ around the candidate track. The second
requirement is that $E/p < 0.3$, where $E$ is the sum of energy
deposited in the calorimeters within a radius $\Delta R = 0.3$ around the candidate track (including
the candidate energy deposit) and $p$ is the candidate track
momentum reconstructed  in the tracker.
The probabilities $P^{\text{on}}(k)$ and $P^{\text{off}}(\mthresh, k)$ are estimated with
single-particle events and thus do not account for the possibility that a long-lived particle might fail the isolation requirements.
In order to accurately model the isolation requirements, the following procedure, which uses generator-level information from a simulation of the BSM model under test, should be used.
The isolation requirements must be determined for each long-lived particle at the generator-level.
The following conditions are imposed:
\begin{equation}
\sum_{j}^{\stackrel{\text{charged particles}}{\Delta R<0.3}} \hspace*{-1em}\pt^{j}  < 50\GeV
\label{eq:GenTkCaloIso}
\quad\text{and}\quad
\sum_{j}^{\stackrel{\text{visible particles}}{\Delta R<0.3}} \hspace*{-1em}E^{j} / p  < 0.3 .
\end{equation}
The equations (\ref{eq:GenTkCaloIso}) represent the generator-level equivalents of the tracker and calorimeter isolation requirements, respectively.
The sum of the tracks \pt around the long-lived candidate is replaced by a sum over transverse momenta of all charged particles around the direction of the long-lived particle.
Similarly, the sum of the calorimeter energy deposits around the long-lived candidate is replaced by a sum over the energies of all visible particles, except the long-lived one, around the direction of the long-lived particle of momentum magnitude $p$.
Muons are not considered in the sum over visible particles since they deposit very little energy in the calorimeter.
If a long-lived particle does not satisfy both isolation requirements expressed by Eqs. (\ref{eq:GenTkCaloIso}), the corresponding factor $P^{\text{off}}(\mthresh, k)$ in Eq.~(\ref{eq:EventAcceptance}) is set to zero.
While pileup effects are not taken into account in the above prescriptions, it was checked that these omissions do not have a significant effect on the results.
The robustness of the fast estimate of the acceptance against the hadronic activity surrounding the long-lived particle is assessed by testing different chargino production mechanisms: direct pair production of charginos, pair production of gluinos that each decay to a heavy quark (b,t) and a chargino, pair production of gluinos that each decay to a light quark (u,d,s,c) and a chargino, and pair production of squarks that each decay to a quark and a chargino.
A $\sim$10\% agreement between the fast estimate of the model acceptance and the full simulation predictions is observed as shown in Fig.~\ref{fig:TkTOFpMSSMValidationIsolation}.
The estimate of the acceptance for chargino pair production is not modified by the isolation requirement since the charginos are always isolated for that production mechanism.

In the two following sections, the fast technique is used to estimate the signal acceptance of the CMS detector for two extensions of the SM.
Given that no significant excess of events over the predicted backgrounds is observed in Ref.~\cite{EXO-12-026},
the signal acceptance estimated with the technique described above is used in calculating cross section limits of these models, at 95\% confidence level (CL).
The limits in this paper are established using the CL$_\mathrm{S}$
approach~\cite{Junk:1999kv, READ:JPG2002} where $p$-values are
computed with a hybrid Bayesian-frequentist
technique~\cite{ATLAS:1379837}.  A log-normal
model~\cite{Eadie, James} is adopted for the nuisance parameters.
The nuisance parameters are: the expected background in the signal region, the integrated
luminosity, and the signal acceptance.
The expected background and the integrated luminosity, as well as their
uncertainties, are provided in Ref.~\cite{EXO-12-026}.
The uncertainty in the signal acceptance is assumed to be 25\% for all signal models.
This value results from adding 10\% to the $\leq$15\% uncertainty reported in Section 6 of Ref.~\cite{EXO-12-026}.
The additional 10\% accounts for the systematic uncertainty incurred by the use of the fast estimation technique to compute the acceptance.
Table~\ref{tab:InfoSummary} summarizes the information needed, in addition to the signal acceptance evaluated with the fast technique, to set limits on a signal model predicting lepton-like charged long-lived particles.

\begin{table*}[htb]
 \caption{
   Summary of the information needed to set limits on a signal model predicting lepton-like charged long-lived particles.
   The mass threshold, the corresponding expected background, and the observed numbers of events, as well as the uncertainty in the signal acceptance evaluated with the fast technique, are provided as a function of the long-lived particle mass.
   \label{tab:InfoSummary}}
  \centering
 \begin{tabular}{c|cccc} \hline
 Mass & $m_{\text{thresh}}$ & Predicted & Data & Signal Unc. \\
 (\GeVns{}) & (\GeVns{})      & backgrounds      & counts     & (\%) \\ \hline
$m < 166$   & $0$   & $44\pm9$       & $42$ & $25$ \\
$166 < m < 330$ & $100$ & $5.6\pm1.1$    & $7$  & $25$ \\
$330 < m < 500$ & $200$ & $0.56\pm0.11$  & $0$  & $25$ \\
$500 < m$    & $300$ & $0.090\pm0.02$ & $0$  & $25$ \\
\hline
 \end{tabular}
\end{table*}

To demonstrate the validity of this procedure, we have used it to translate the estimate of the signal acceptance to the 95\% CL limit on the production cross section for two of the models considered in  Ref.~\cite{EXO-12-026}.
The limits obtained on the pair production and inclusive production of staus in the context of the gauge mediated supersymmetry breaking (GMSB) model
are shown in Table~\ref{tab:TkTOFValidationPlot3} for signal acceptance values estimated with full simulation and with the procedure described above.
The limits obtained with the two techniques agree within 8\%.  The differences are due only to a small difference in the signal acceptance computed with the two different
techniques and to the larger uncertainty assigned to the acceptance from the fast technique.

The results in Table 1 may be compared with those shown in Table 7 and Fig. 8 of Ref.~\cite{EXO-12-026}.
In making the comparisons, it should be noted that there are important differences in the way that \mthresh is chosen in the two cases.
In Ref.~\cite{EXO-12-026} the value was varied in steps of 10\GeV in order to optimise the resultant limit.
In this paper a simpler approach has been followed with \mthresh varied in 100\GeV steps and chosen to be the largest value satisfying the condition  $\mthresh \leq 0.6 m$.
The latter approach generally results in a somewhat higher estimation of the background and therefore a more conservative limit.
In the most extreme case, the pair production of staus, with $m = 308\GeV$, the background is estimated to be $5.6\pm1.1$ events with  $\mthresh = 100\GeV$, compared with the estimate of $0.7\pm0.1$ events obtained with  $\mthresh = 190$\GeV in Ref.~\cite{EXO-12-026},
resulting in a cross section limit that is about three times higher in the former case.
Nonetheless, the limits agree within $\sim$15\% in almost all cases, allowing restrictive limits to be set on a general class of models.

\begin{table*}[htb]
 \topcaption{
   Signal acceptance estimated from the fast technique and with the full simulation of the detector, as well as the corresponding expected and observed cross section limits.
   Results are provided for both the pair production and the inclusive production of staus as predicted by the GMSB model.
   The mass threshold, the corresponding expected background and the observed numbers of events is also shown.
   \label{tab:TkTOFValidationPlot3}}
  \centering
 \resizebox{\textwidth}{!}{
 \begin{tabular}{c|c|cc|ccc|ccc} \hline
 Mass & \mthresh & Predicted & Data & \multicolumn{3}{c|}{Full simulation} &  \multicolumn{3}{c}{Fast technique} \\
 (\GeVns{}) & (\GeVns{})      & backgrounds      & counts     & Acc. & Exp. (fb) & Obs. (fb)                    & Acc. & Exp. (fb) & Obs. (fb) \\
\hline \multicolumn{10}{c}{Pair production of staus} \\ \hline
 126 &   0 & $ 44   \pm  9   $ &  42 &     0.24 &  4.38 &  4.11      &  0.24 &  4.53 &  4.24 \\
 156 &   0 & $ 44   \pm  9   $ &  42 &     0.28 &  3.66 &  3.43      &  0.29 &  3.81 &  3.57 \\
 200 & 100 & $  5.6 \pm  1.1 $ &   7 &     0.34 &  1.06 &  1.28      &  0.35 &  1.08 &  1.30 \\
 247 & 100 & $  5.6 \pm  1.1 $ &   7 &     0.40 &  0.90 &  1.09      &  0.40 &  0.93 &  1.13 \\
 308 & 100 & $  5.6 \pm  1.1 $ &   7 &     0.46 &  0.77 &  0.93      &  0.47 &  0.79 &  0.96 \\
 370 & 200 & $  0.56\pm  0.11$ &   0 &     0.53 &  0.41 &  0.31      &  0.53 &  0.42 &  0.32 \\
 494 & 200 & $  0.56\pm  0.11$ &   0 &     0.61 &  0.36 &  0.27      &  0.62 &  0.37 &  0.28 \\
 745 & 300 & $  0.09\pm  0.02$ &   0 &     0.66 &  0.24 &  0.24      &  0.67 &  0.25 &  0.24 \\
1029 & 300 & $  0.09\pm  0.02$ &   0 &     0.58 &  0.28 &  0.27      &  0.59 &  0.28 &  0.27 \\
\hline \multicolumn{10}{c}{Inclusive production of staus} \\ \hline
 126 &   0 & $ 44   \pm  9   $ &  42 &     0.25 &  4.22 &  3.95      &  0.25 &  4.43 &  4.15 \\
 156 &   0 & $ 44   \pm  9   $ &  42 &     0.32 &  3.21 &  3.01      &  0.32 &  3.38 &  3.16 \\
 200 & 100 & $  5.6 \pm  1.1 $ &   7 &     0.41 &  0.87 &  1.05      &  0.42 &  0.90 &  1.09 \\
 247 & 100 & $  5.6 \pm  1.1 $ &   7 &     0.50 &  0.72 &  0.87      &  0.50 &  0.76 &  0.91 \\
 308 & 100 & $  5.6 \pm  1.1 $ &   7 &     0.56 &  0.64 &  0.77      &  0.56 &  0.67 &  0.81 \\
 370 & 200 & $  0.56\pm  0.11$ &   0 &     0.60 &  0.36 &  0.27      &  0.60 &  0.37 &  0.28 \\
 494 & 200 & $  0.56\pm  0.11$ &   0 &     0.66 &  0.33 &  0.25      &  0.65 &  0.35 &  0.26 \\
 745 & 300 & $  0.09\pm  0.02$ &   0 &     0.67 &  0.24 &  0.23      &  0.67 &  0.25 &  0.24 \\
1029 & 300 & $  0.09\pm  0.02$ &   0 &     0.58 &  0.28 &  0.27      &  0.58 &  0.29 &  0.28 \\
\hline
 \end{tabular}}
\end{table*}

Because the online selection in Ref.~\cite{EXO-12-026} uses a missing transverse energy (\MET)
trigger in combination with a single-muon trigger, there is one caveat to the proposed factorization method: the efficiency of the \MET trigger cannot be modeled accurately in terms of single
long-lived particle kinematic properties.
Accounting for the presence of other undetectable particles using a Monte Carlo method would not help because \MET
often depends significantly on detector effects due to the other particles.
The assumption that the \MET trigger adds negligibly to the event
selection performed by the muon trigger must therefore be satisfied
in order to apply the method to a given signal.  Deviations from
this assumption would result in an underestimation of the signal acceptance.
The assumption is satisfied by models with lepton-like long-lived particles.
Models with long-lived colored particles, such as top squarks
or gluinos, do not satisfy this condition and thus cannot currently be tested with the technique presented in this paper.
Long-lived colored particles hadronize in color singlet bound states~\cite{Fairbairn:2006gg} that could interact with the detector material leading to complex situations, detailed in Refs.~\cite{Chatrchyan:2012sp, EXO-12-026}, and inducing significant instrumental \MET.
For instance, pair-produced colored long-lived particles may hadronize to a charged and a neutral hadron.
In this case the \MET is strongly modified by the presence of the neutral hadron since it is not visible in the tracker and deposits only ${\mathcal{O}}(1)$\GeV in the calorimeter.
Moreover, the interactions of the color singlet bound states with matter may lead to a modification of the electric charge of the bound states~\cite{Fairbairn:2006gg}.  In this case, the hadron containing a long-lived colored particle can be electrically charged at production, but neutral in the muon system and therefore fail the muon reconstruction.
The probabilities $P^{\text{on}}(k)$ and $P^{\text{off}}(\mthresh, k)$ do not account for the possible modification of the electric charge experienced by the hadron-like particles.

\section{Constraints on the pMSSM}
\label{sec:pmssm}

The pMSSM is a 19-parameter realization of the minimal supersymmetric model (MSSM)~\cite{Djouadi:1998di} that captures most of the
phenomenological features of the R-parity conserving MSSM.
The free parameters of the pMSSM, in addition to the SM parameters, are: (1) the gaugino mass parameters $M_1$, $M_2$, and $M_3$;
(2) the ratio of the vacuum expectation values of the two Higgs doublets $\tan\beta=v_2/v_1$;
(3) the higgsino mass parameter $\mu$ and the pseudoscalar Higgs mass $m_A$;
(4) the 10 sfermion mass parameters $m_{\widetilde{F}}$, where $\widetilde{F} = \widetilde{Q}_1, \widetilde{U}_1, \widetilde{D}_1,  \widetilde{L}_1, \widetilde{E}_1, \widetilde{Q}_3, \widetilde{U}_3, \widetilde{D}_3,  \widetilde{L}_3, \widetilde{E}_3$ (imposing degeneracy of the first two generations $m_{\widetilde{Q}_1}\equiv m_{\widetilde{Q}_2}$,  $m_{\widetilde{L}_1}\equiv m_{\widetilde{L}_2}$, $\dots$); and 5) the trilinear couplings $A_\PQt$, $A_\PQb$, and $A_\tau$.
To minimize the theoretical uncertainties in the Higgs sector, these parameters are defined
at the electroweak symmetry breaking scale.

In the pMSSM, all MSSM parameters are allowed to vary freely, subject to the requirement that the
model is consistent with some basic constraints:
first, the sparticle spectrum must be free of tachyons and cannot lead to color or charge breaking minima in the scalar potential.
We also require that the model is consistent with electroweak symmetry breaking and that the Higgs potential is bounded from below.
Finally, in this study, we also require the lightest SUSY particle to be the lightest neutralino~(\PSGczDo).
Furthermore, for practical reasons, we limit our study to the pMSSM sub-space chosen to cover sparticle masses up to about 3\TeV.
Table~\ref{tab:pmssm} summarizes the boundaries of the considered sub-space.

\begin{table}[h!]
 \centering
   \topcaption{\label{tab:sub-space} The pMSSM parameter space used in the scan.}
   \label{tab:pmssm}
   \begin{tabular}{|rcl|} \hline
        $ -3 \le $&$ M_1, M_2 $&$\le 3\TeV$\\
        $ 0 \le $&$ M_3 $&$\le 3\TeV$\\
        $ -3 \le$&$ \mu $&$\le 3\TeV$\\
        $ 0 \le $&$ m_A $&$\le 3\TeV$\\
        $ 2 \le $&$ \tan\beta $&$\le$60\\
        $ 0 \le $&$ \widetilde{Q}_{1,2}, \widetilde{Q}_3 $&$ \le 3\TeV$ \\
        $ 0 \le $&$ \widetilde{U}_{1,2}, \widetilde{U}_3 $&$ \le 3\TeV$ \\
        $ 0 \le $&$ \widetilde{D}_{1,2}, \widetilde{D}_3 $&$ \le 3\TeV$ \\
        $ 0 \le $&$ \widetilde{L}_{1,2}, \widetilde{L}_3 $&$ \le 3\TeV$ \\
        $ 0 \le $&$ \widetilde{E}_{1,2}, \widetilde{E}_3 $&$ \le 3\TeV$ \\
        $ -7 \le $&$ A_t, A_b, A_\tau $&$\le 7\TeV$ \\
        \hline
   \end{tabular}
\end{table}

In Ref.~\cite{Sekmen:2011cz}, some previous CMS published results were reinterpreted in the context of this pMSSM sub-space.  This analysis, however, did not consider the region of parameter space in which long-lived charginos are predicted.
Using the technique described in Section~\ref{sec:techniqueIntro}, we can extend the results of  Ref.~\cite{Sekmen:2011cz} to regions of the parameter space leading to long-lived particles.

We sample 20 million points in a pMSSM parameter space
from a prior probability density function that encodes results from indirect SUSY searches and pre-LHC searches as done in Ref.~\cite{Sekmen:2011cz}.
From this set we select 7205 points with a Higgs boson mass in the range $120 \le m_{\Ph} \le 130$\GeV and predicting long-lived ($c\tau > 50$\unit{cm}) charginos.
Tightening further the mass window to $123 \le m_{\Ph} \le 128$\GeV to reflect more recent constraints from the Higgs boson mass measurements~\cite{Chatrchyan:2013mxa,Khachatryan:2014ira,Aad:2012tfa} reduces the size of the subset by 45\% but does not further constrain the chargino mass or lifetime.
We therefore use the $120 \le m_{\Ph} \le 130$\GeV window in order to minimize the statistical uncertainty.

For each of the 7205 points in this subset, we have generated 10\,000 events using \PYTHIA v6.426~\cite{Sjostrand:2006za}.
Both direct pair production of charginos and indirect chargino production through the decay of heavier SUSY particles were considered for this study.
The generated events have been used to evaluate the signal acceptance of the HSCP search, given the chargino kinematic properties predicted by \PYTHIA for the considered pMSSM sub-space.

The fast technique is used to obtain acceptance values expected to be
in good agreement with the full simulation prediction.
The predicted signal acceptance is then used to
compute 95\% CL limits on the 7205 analysed pMSSM parameter points.
A parameter point is excluded if the observed limit obtained on the cross section is less than the
theoretical prediction at leading-order as calculated by \PYTHIA.
The use of leading-order instead of next-to-leading-order theoretical cross section is driven by practical considerations given the large number of parameter points considered.

\begin{figure}[tbhp]
 \begin{center}
  \includegraphics[width=0.49\textwidth]{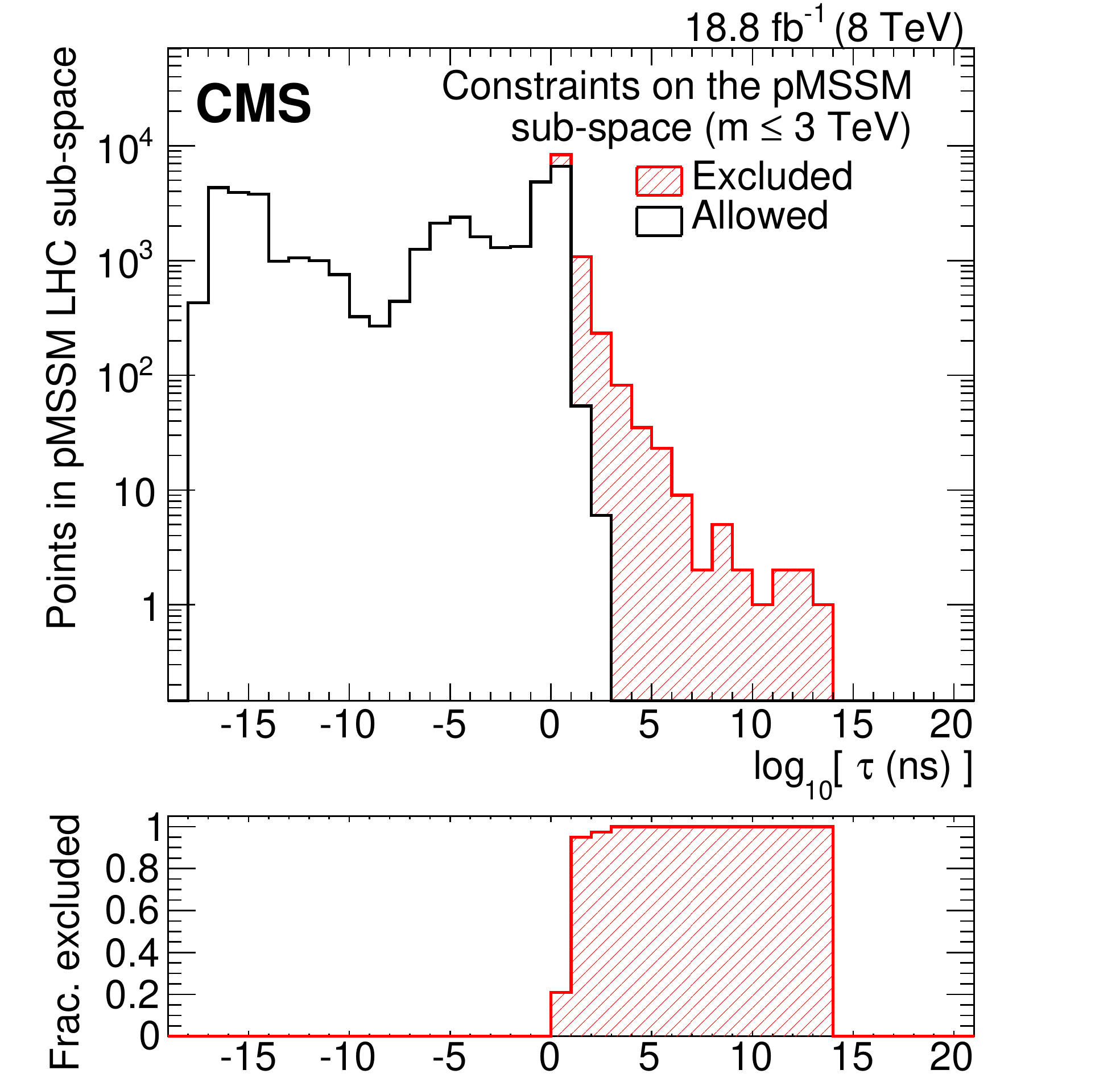}
  \includegraphics[width=0.49\textwidth]{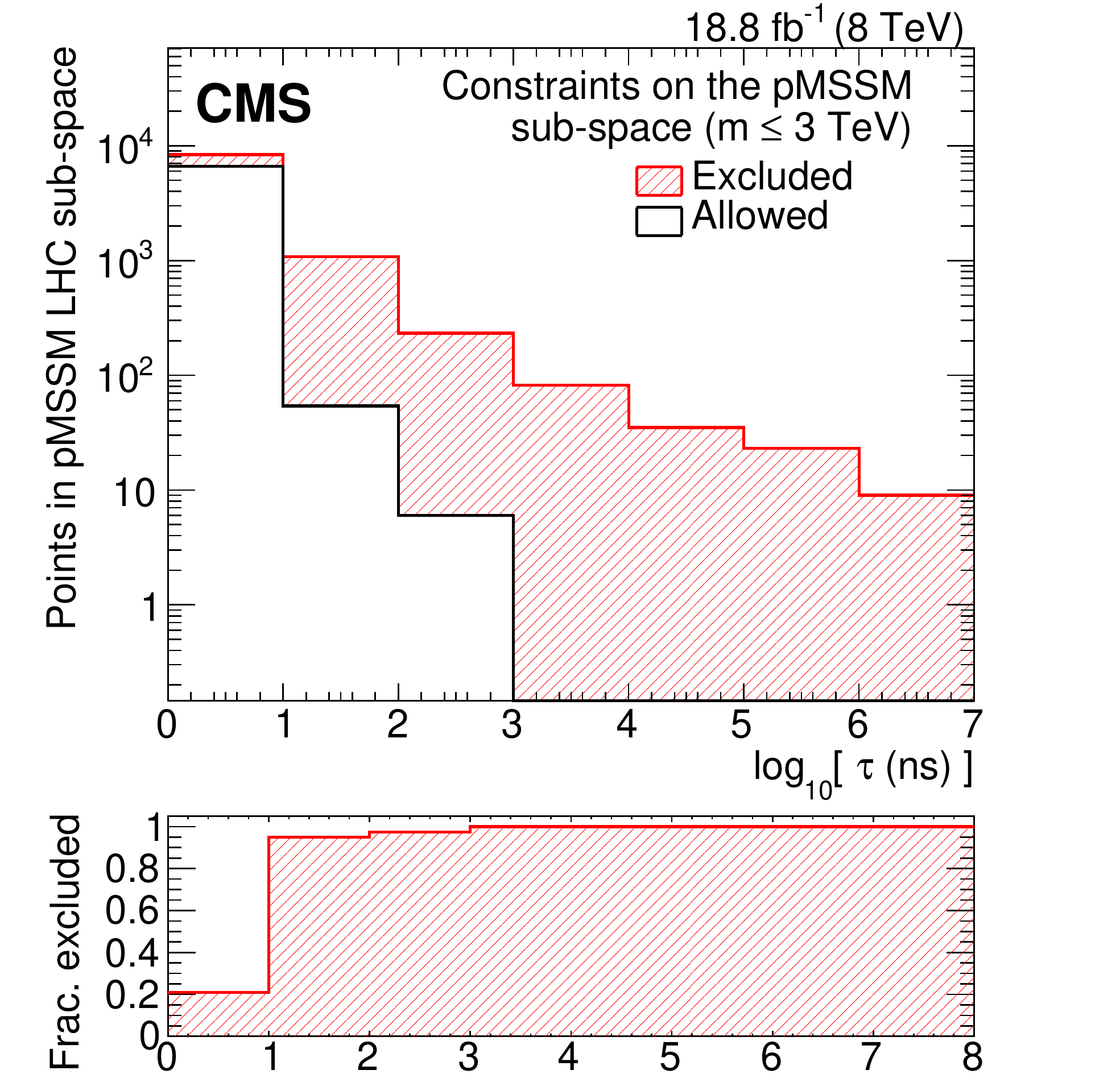}
 \end{center}
 \caption{(\cmsLeft)  Number of pMSSM points, in the sub-space covering sparticle masses up to about 3\TeV,  that are excluded at a 95\% CL (hatched red) or allowed (white) as a function of the chargino lifetime.  (\cmsRight)  Enlargement of the long-lived region.
    The bottom panel shows the fraction of pMSSM points excluded by the analysis based on the results from the HSCP search~\cite{EXO-12-026}.
    \label{fig:TkTOFpMSSMLimits}}
\end{figure}

Figure~\ref{fig:TkTOFpMSSMLimits} shows the fraction of
parameter points excluded as a function of the chargino lifetime.
The fraction of excluded model points with a chargino lifetime longer than 1000\unit{ns} (10\unit{ns}) is 100.0\% (95.9\%).
Although these values depend on the random point sampling in the 19-dimensional pMSSM parameter space, it is remarkable that a high fraction of the points predicting long-lived charginos are excluded.

\begin{figure*}[tbhp]
 \begin{center}
  \includegraphics[width=0.49\textwidth]{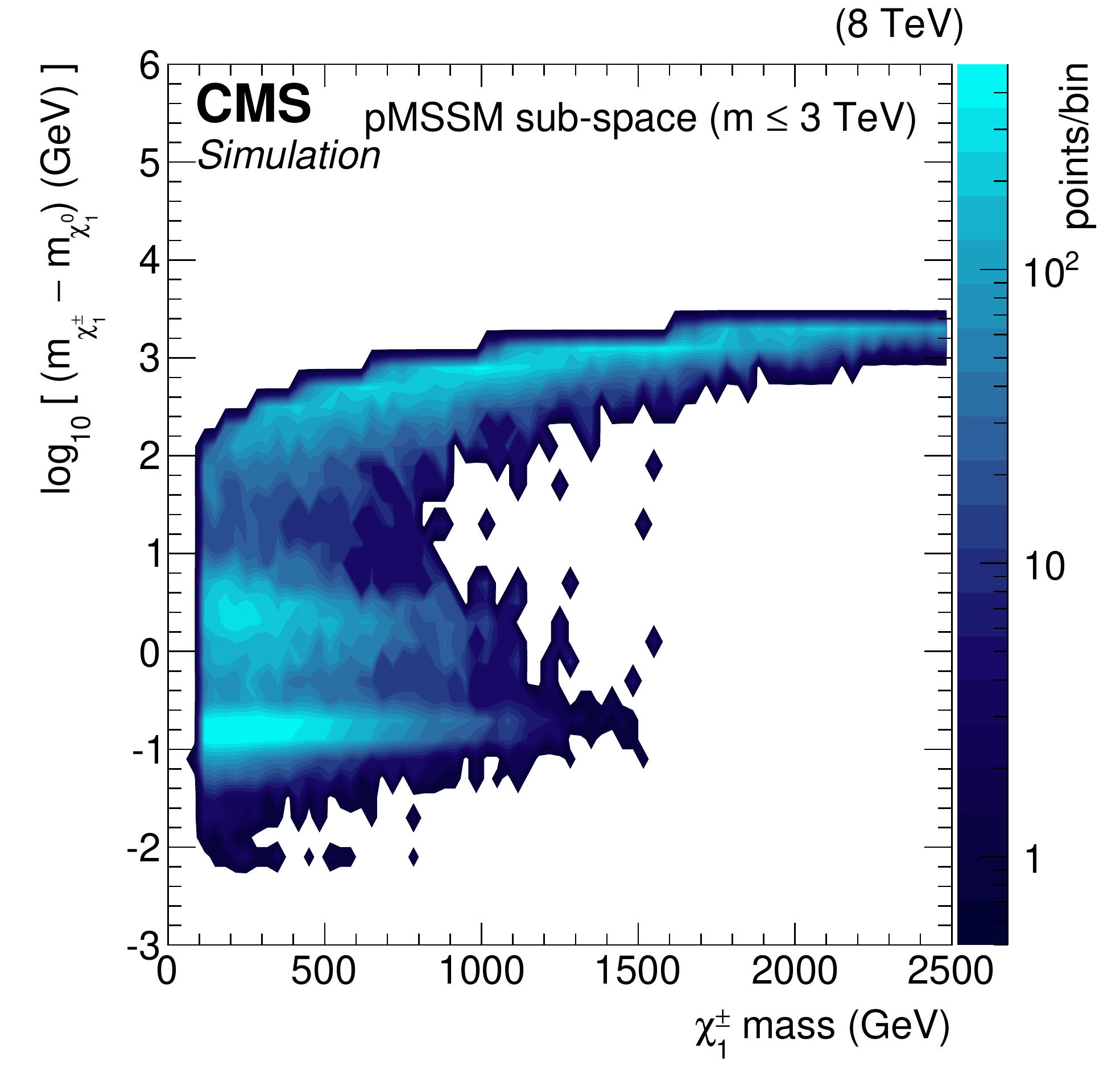}
  \includegraphics[width=0.49\textwidth]{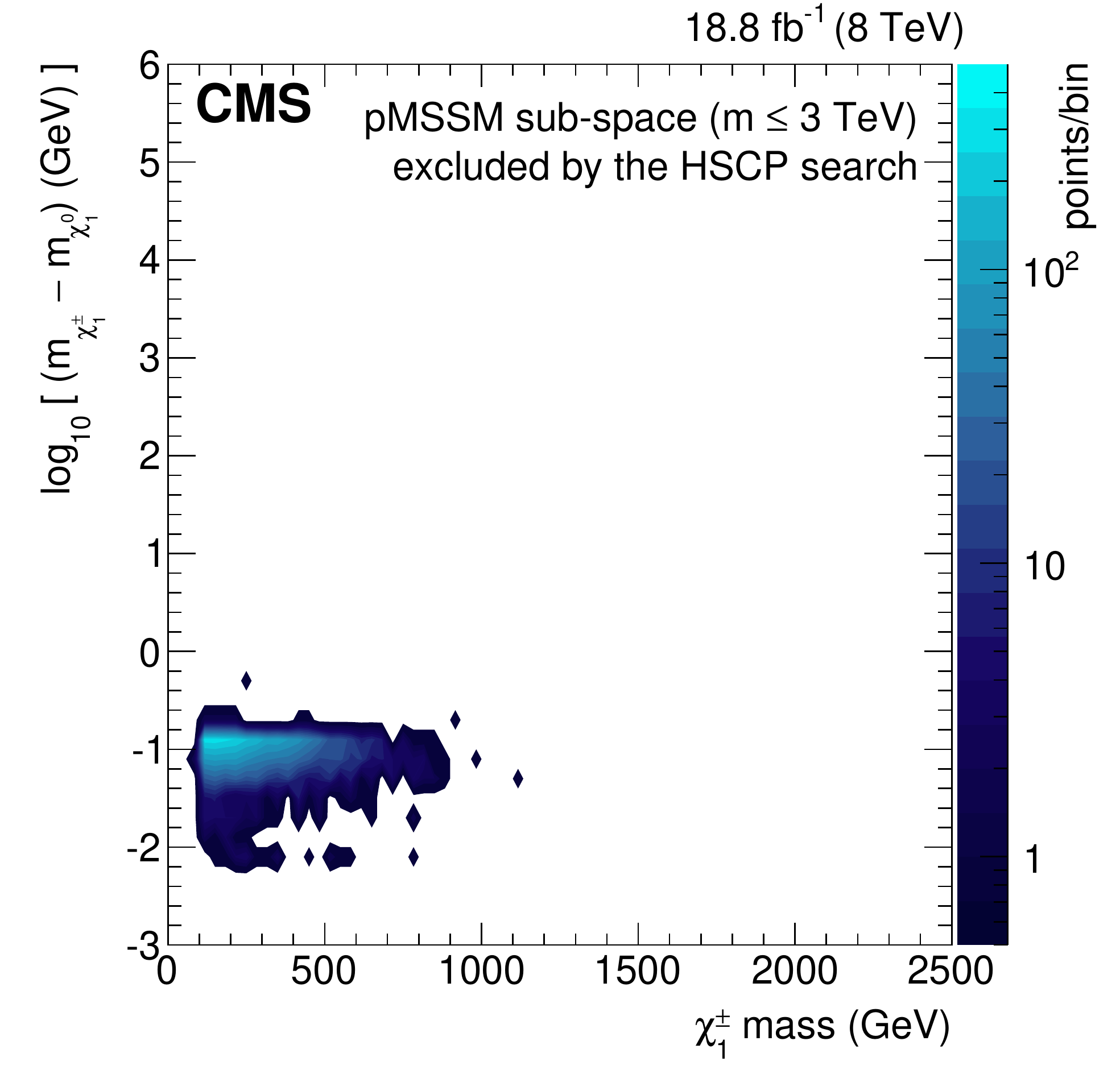} \\
  \includegraphics[width=0.49\textwidth]{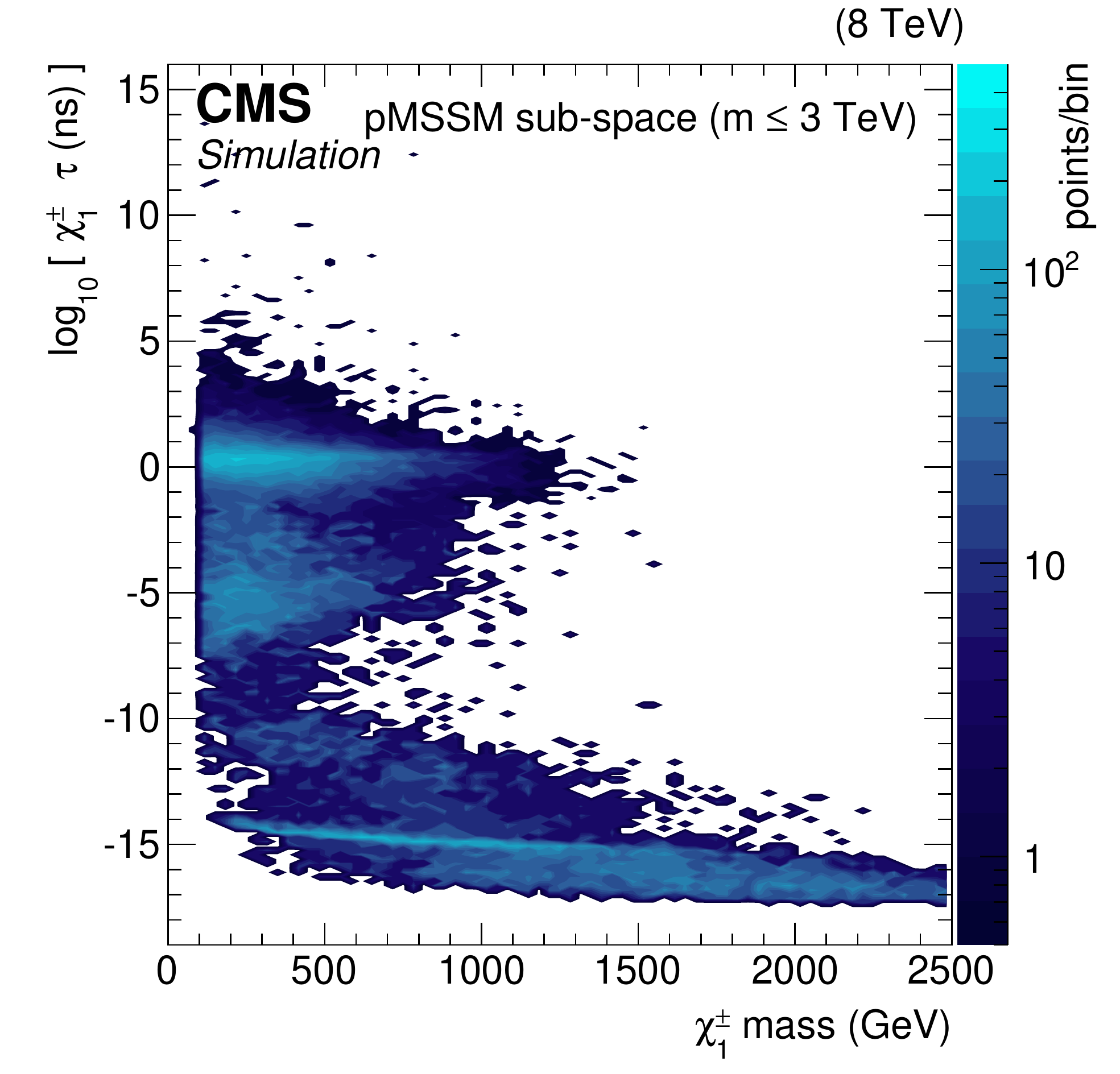}
  \includegraphics[width=0.49\textwidth]{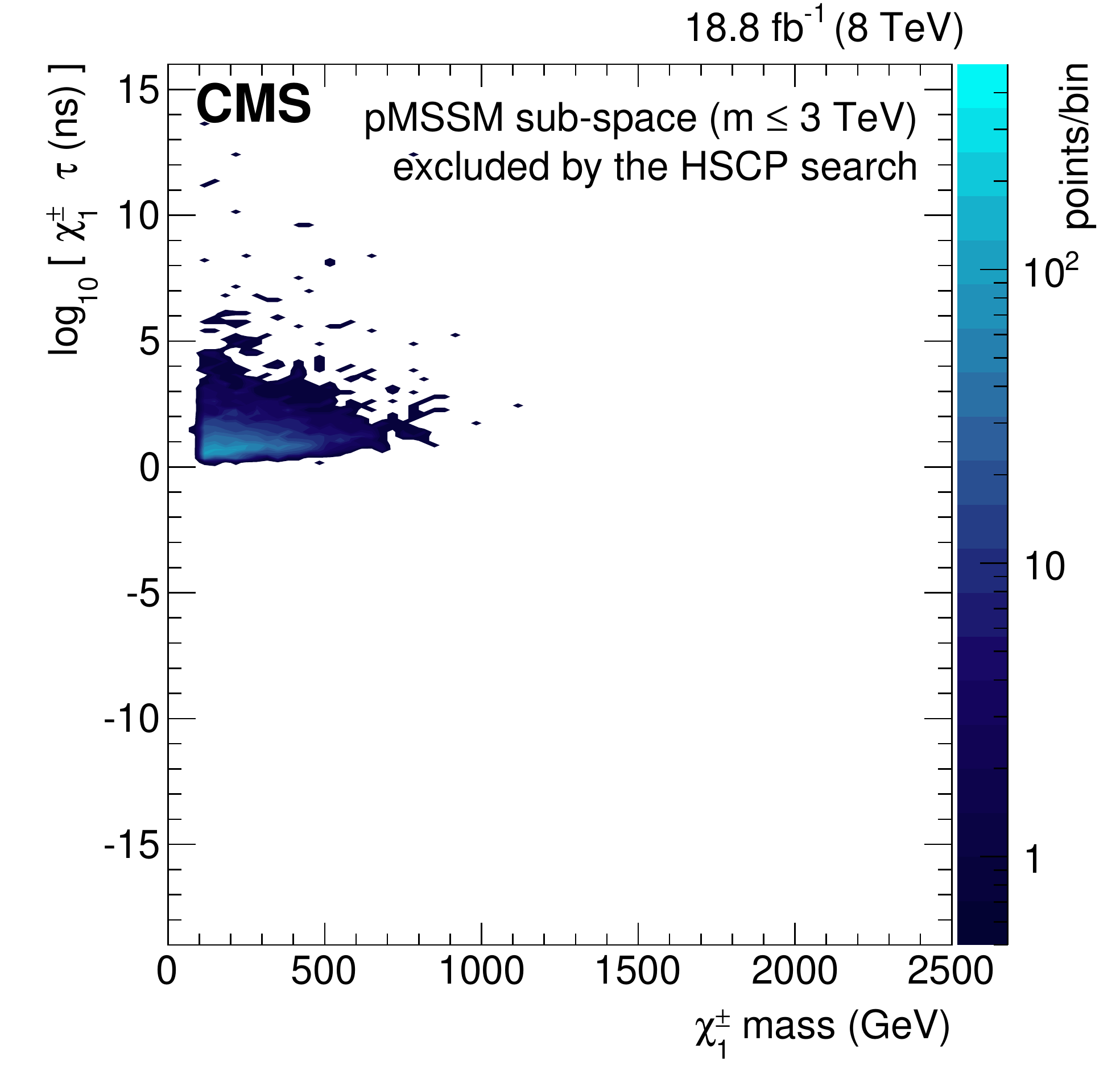} \\
\end{center}
 \caption{Number of pMSSM parameter points in the sub-space covering sparticle masses up to about 3\TeV shown as a function of the chargino mass and (upper row) of the mass difference between the chargino and the neutralino, and (lower row) chargino lifetime.  The left panels show the entire set of points considered while the right panels show the set of points excluded by the analysis based on the results from the HSCP search~\cite{EXO-12-026}.
    \label{fig:pMSSM_MM}}
\end{figure*}

Figure~\ref{fig:pMSSM_MM} shows the number
of parameter points predicted and excluded
by the analysis of the results obtained in Ref.~\cite{EXO-12-026} as a
function of the chargino mass, chargino lifetime, and the mass difference between the chargino and the neutralino.

\section{Constraints on the AMSB model}
\label{sec:amsb}

In the AMSB model~\cite{Chen:1996ap, Giudice1998, Randall1999} the lightest chargino and neutralino are almost degenerate  ($m_{\PSGcpm_1} - m_{\PSGczDo} \leq 1$\GeV), where the neutralino is the lightest SUSY particle.
In this model, the chargino lifetime, expected to be of the order of a nanosecond or larger, is determined by the mass splitting with the neutralino.

Previous searches for AMSB charginos~\cite{Aad:2013yna,CMS:2014gxa} looked for a chargino decaying within the tracking volume into a neutralino and a soft charged pion.
The pion has a momentum of $\sim100$ MeV and is generally not reconstructed. The experimental chargino signature therefore takes the form of a disappearing track inside the tracking system.
The main limitation of that search technique is that the sensitivity drops quickly as the lifetime increases because the chargino is required to decay within the tracking region.

In contrast, the sensitivity of the search for HSCP~\cite{EXO-12-026} is maximal when the chargino decay occurs outside the detector.  These two searches are therefore complementary.
In this context, the fast technique discussed in the previous sections can be used to assess the limits set by Ref.~\cite{EXO-12-026} on long-lived charginos in the AMSB model.

The minimal version of the AMSB model is fully characterized by four parameters:
the ratio of Higgs doublet vacuum expectation values at the electroweak scale,
the sign of the higgsino mass term, the universal scalar mass ($m_0$),
and the gravitino mass ($m_{3/2}$), which dictates the value of the chargino mass.
The values of the first two parameters are set to $\tan{\beta}=5$ and $\mu>0$.
The scalar mass is set to a large value (1\TeV) in order to prevent the appearance of tachyonic sleptons.
Gravitino masses ranging from 3.5 to 32\TeV are used in order to scan chargino masses from 100 to 900\GeV.

\begin{figure}[tbhp]
 \begin{center}
  \includegraphics[width=0.49\textwidth]{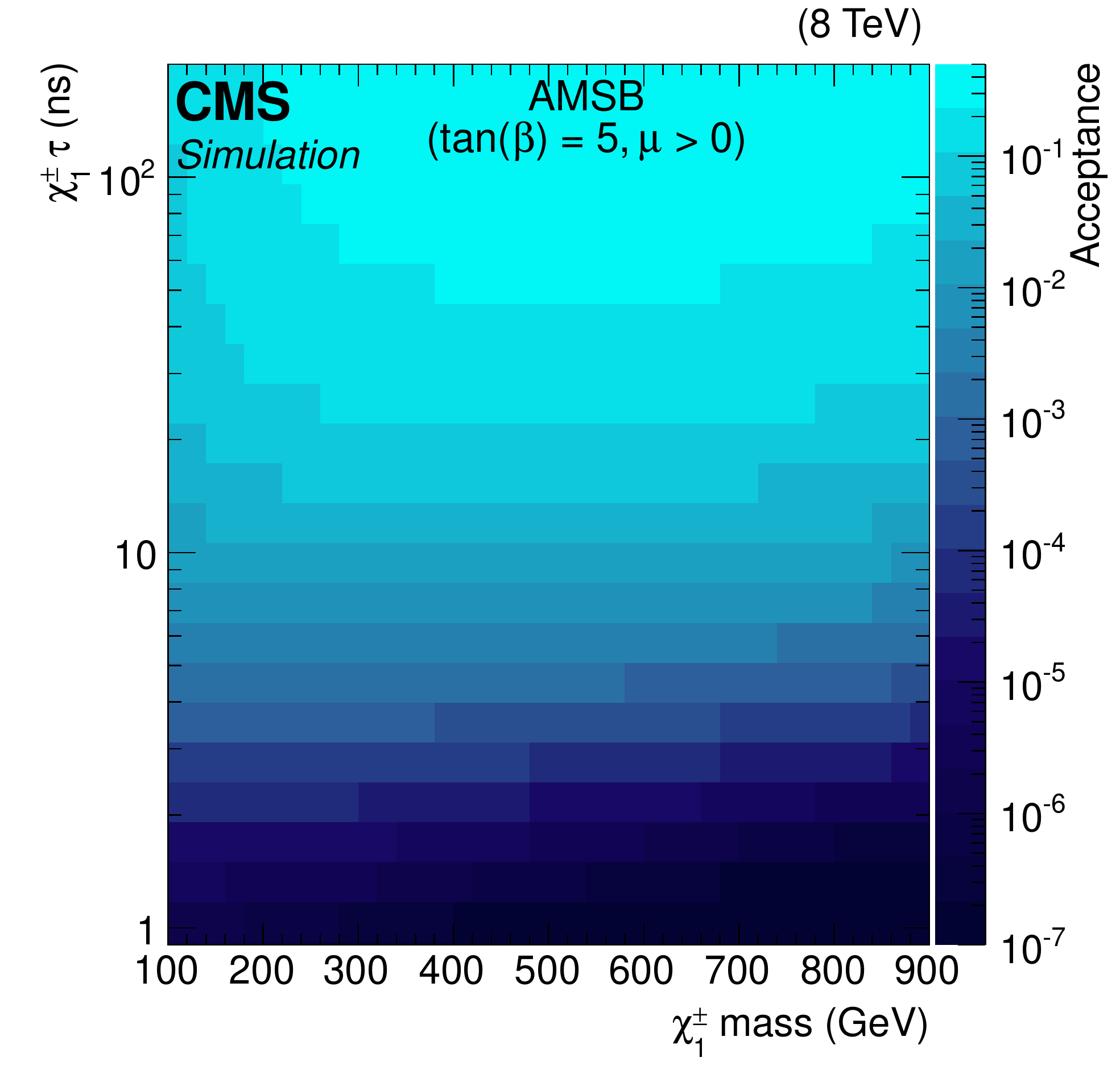}
  \includegraphics[width=0.49\textwidth]{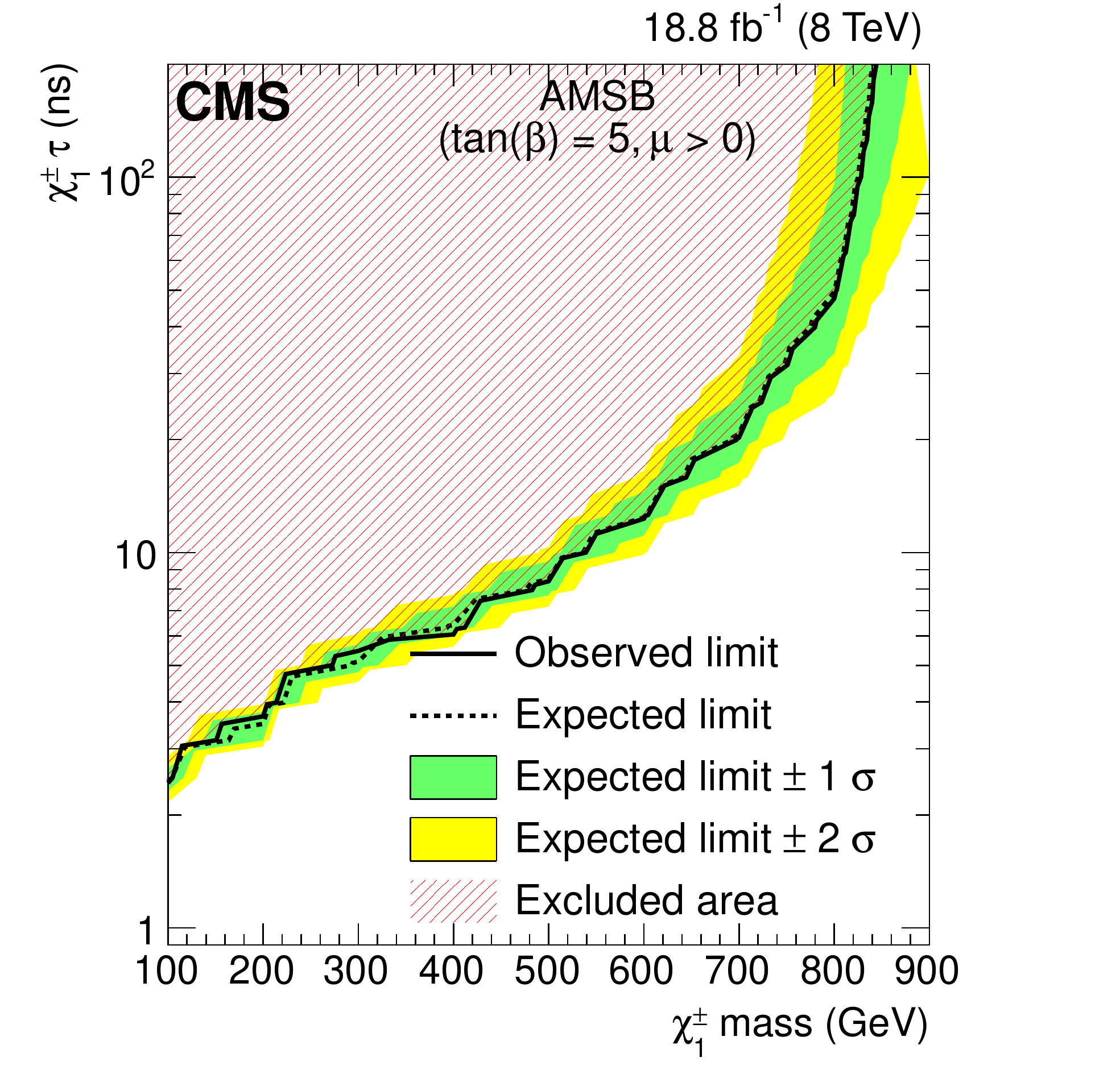}
 \end{center}
 \caption{(\cmsLeft) Signal acceptance as a function of chargino mass for the AMSB model as predicted by the fast technique.  (\cmsRight) Observed and expected excluded region on the chargino mass and lifetime parameter space in the context of the AMSB model with  $\tan\beta = 5$ and $\mu \ge 0$.  The excluded region is indicated by the hatched area.
    \label{fig:TkTOFAMSBLimit}}
\end{figure}

Samples of simulated charginos with masses from 100 to 900\GeV and lifetimes from 1\unit{ns} to 10\unit{$\mu$s} are produced with \PYTHIA v6.426.  The SUSY mass spectrum and the decay tables are calculated using \ISASUSY with \ISAJET v7.80~\cite{Isajet}.
For each sample, 10\,000 events are generated using an inclusive SUSY production and the acceptance of the search for long-lived particles is estimated using the technique described in Section~\ref{sec:techniqueIntro}.
The estimated signal acceptance is shown in Fig.~\ref{fig:TkTOFAMSBLimit} (\cmsLeft). The acceptance is reduced for short lifetimes because the probability that a particle reaches the muon system before decaying, is exponentially smaller.
As explained in Section \ref{sec:techniqueIntro}, a systematic uncertainty of 25\% is assigned to the signal acceptance.
A point in the mass--lifetime parameter space of the AMSB model is considered to be excluded when the 95\% CL observed limit on the cross section is lower than the leading-order theoretical cross section.
The excluded region in this plane is shown in Fig.~\ref{fig:TkTOFAMSBLimit} (\cmsRight).
These results extend those from previous searches at LHC experiments~\cite{Aad:2013yna,CMS:2014gxa} by excluding charginos with lifetimes $\gtrsim$100\unit{ns} up to masses of about 800\GeV.
While the signal acceptance remains nearly constant over a wide mass range, heavier charginos cannot be excluded because of their smaller production cross section.
The sensitivity of the search for HSCP~\cite{EXO-12-026} is limited to charginos with an average lifetime $\geq 3$\unit{ns} while previous searches based on short track signatures are sensitive to lifetimes of $\sim 0.1$\unit{ns}~\cite{Aad:2013yna,CMS:2014gxa}.

\section{Summary}

The results of the CMS search for long-lived charged particles have been analysed to set constraints on the phenomenological minimal supersymmetric standard model (pMSSM)
and on the anomaly-mediated SUSY breaking model (AMSB), both of which predict the existence of long-lived massive particles in certain regions of their parameter space.
A novel technique for estimating the signal acceptance with an accuracy of 10\% is presented.  This technique only uses generator-level  information, while the integrated luminosity, the expected standard model background yields, and the corresponding uncertainties are taken from a previous CMS search~\cite{EXO-12-026}.
The technique and the tabulated probabilities, available as supplementary material to this paper~\cite{ProbMaps}, can be used by others to estimate the CMS exclusion limits for different models predicting long-lived lepton-like particles.
In the context of the AMSB model, charginos with lifetimes $\gtrsim 100$\unit{ns} (3\unit{ns}) and masses up to about 800\GeV (100\GeV) are excluded at 95\% confidence level.
The most stringent limits to date are set on the long-lived sector of the pMSSM  sub-space that covers SUSY particle masses up to about 3\TeV.
In this sub-space, 95.9\% (100\%) of the points with a chargino lifetime $\tau \geq$ 10\unit{ns} (1000\unit{ns}) are excluded by the present analysis of the results from the CMS search in Ref.~\cite{EXO-12-026}.
These are the first constraints on the pMSSM obtained at the LHC.

\begin{acknowledgments}
We congratulate our colleagues in the CERN accelerator departments for the excellent performance of the LHC and thank the technical and administrative staffs at CERN and at other CMS institutes for their contributions to the success of the CMS effort. In addition, we gratefully acknowledge the computing centres and personnel of the Worldwide LHC Computing Grid for delivering so effectively the computing infrastructure essential to our analyses. Finally, we acknowledge the enduring support for the construction and operation of the LHC and the CMS detector provided by the following funding agencies: BMWFW and FWF (Austria); FNRS and FWO (Belgium); CNPq, CAPES, FAPERJ, and FAPESP (Brazil); MES (Bulgaria); CERN; CAS, MoST, and NSFC (China); COLCIENCIAS (Colombia); MSES and CSF (Croatia); RPF (Cyprus); MoER, ERC IUT and ERDF (Estonia); Academy of Finland, MEC, and HIP (Finland); CEA and CNRS/IN2P3 (France); BMBF, DFG, and HGF (Germany); GSRT (Greece); OTKA and NIH (Hungary); DAE and DST (India); IPM (Iran); SFI (Ireland); INFN (Italy); MSIP and NRF (Republic of Korea); LAS (Lithuania); MOE and UM (Malaysia); CINVESTAV, CONACYT, SEP, and UASLP-FAI (Mexico); MBIE (New Zealand); PAEC (Pakistan); MSHE and NSC (Poland); FCT (Portugal); JINR (Dubna); MON, RosAtom, RAS and RFBR (Russia); MESTD (Serbia); SEIDI and CPAN (Spain); Swiss Funding Agencies (Switzerland); MST (Taipei); ThEPCenter, IPST, STAR and NSTDA (Thailand); TUBITAK and TAEK (Turkey); NASU and SFFR (Ukraine); STFC (United Kingdom); DOE and NSF (USA).

Individuals have received support from the Marie-Curie programme and the European Research Council and EPLANET (European Union); the Leventis Foundation; the A. P. Sloan Foundation; the Alexander von Humboldt Foundation; the Belgian Federal Science Policy Office; the Fonds pour la Formation \`a la Recherche dans l'Industrie et dans l'Agriculture (FRIA-Belgium); the Agentschap voor Innovatie door Wetenschap en Technologie (IWT-Belgium); the Ministry of Education, Youth and Sports (MEYS) of the Czech Republic; the Council of Science and Industrial Research, India; the HOMING PLUS programme of Foundation for Polish Science, cofinanced from European Union, Regional Development Fund; the Compagnia di San Paolo (Torino); the Consorzio per la Fisica (Trieste); MIUR project 20108T4XTM (Italy); the Thalis and Aristeia programmes cofinanced by EU-ESF and the Greek NSRF; and the National Priorities Research Program by Qatar National Research Fund.
\end{acknowledgments}

\bibliography{auto_generated}
\appendix
\numberwithin{figure}{section}
\section{Details of the fast technique}
\label{sec:appendix}

The probabilities $P^{\text{on}}(k)$ and $P^{\text{off}}(\mthresh, k)$, introduced in Section~\ref{sec:techniqueIntro}, are
computed  using samples of single long-lived particles, uniformly distributed in $\eta$ and $\beta$, and produced in \PYTHIA\ v6.426~\cite{Sjostrand:2006za}.
Twenty samples, each containing one million stau particles, were produced for the following long-lived particle masses: 100, 126, 156,
200, 247, 308, 370, 432, 494, 557, 651, 745, 871, 1029, 1200, 1400,
1600, 1800, 2000, and 2500\GeV.
These generated events were then processed by the full CMS simulation and reconstruction software.  The full simulation includes pileup effects.
The selection requirements adopted by the ``tracker+TOF" analysis described in Ref.~\cite{EXO-12-026} were then applied to the reconstructed events in order to evaluate the probability that a particle with kinematics $k$ can pass the selection criteria.

\begin{figure*}[tbh]
 \begin{center}
  \includegraphics[width=0.90\textwidth]{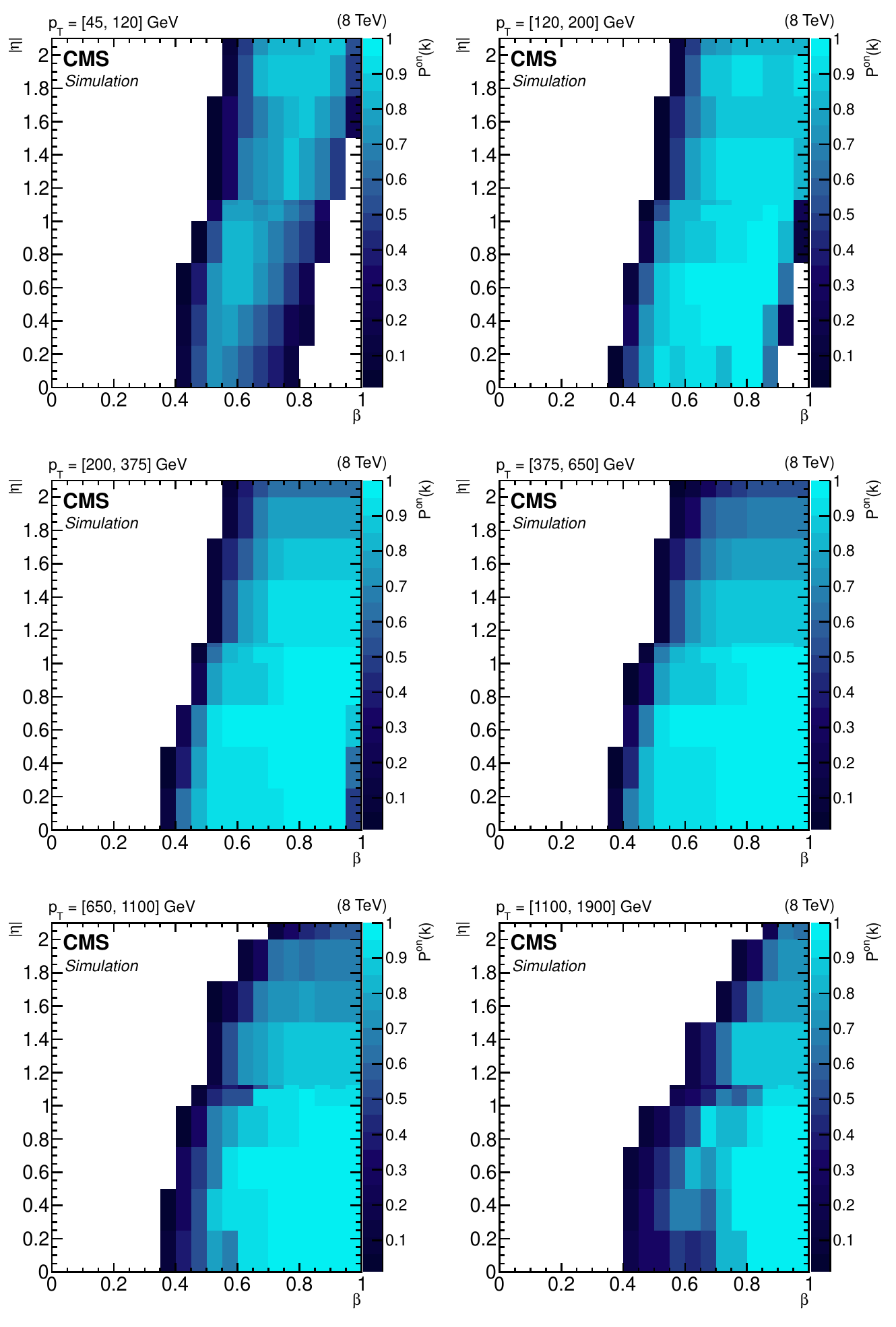}
 \end{center}
 \caption{Values taken by the probability $P^{\text{on}}(k)$ as a function of the true particle-variables $\pt$, $\beta$, and $\abs{\eta}$.
The binning in \pt in these figures is coarser than the one used in the study. }
    \label{fig:TkTOFTrigEff}
\end{figure*}

\begin{figure*}[tbh]
 \begin{center}
  \includegraphics[width=0.90\textwidth]{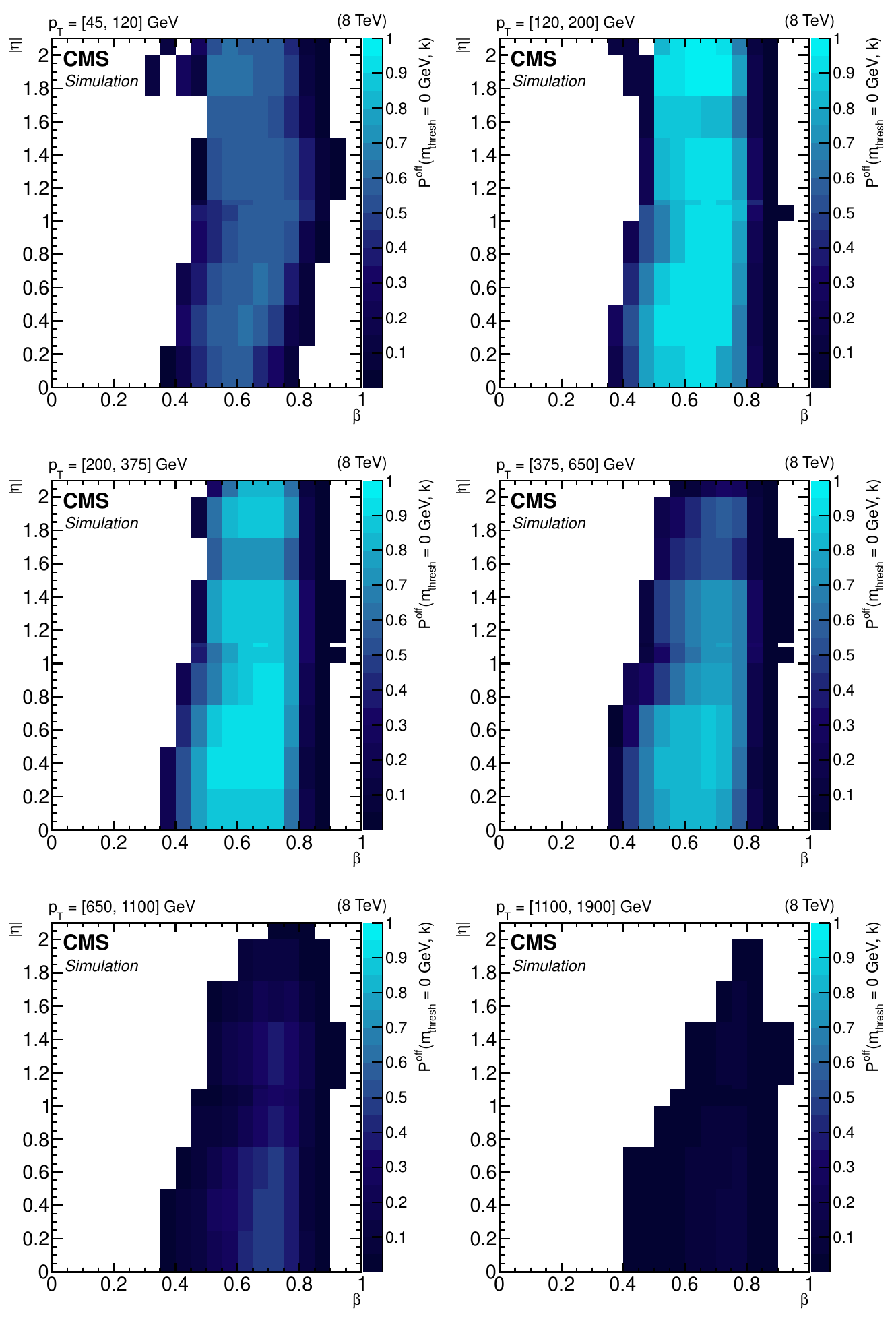}
 \end{center}
 \caption{Values taken by the probability $P^{\text{off}}(0, k)$ as a function of
the true particle-variables $\pt$, $\beta$, and $\abs{\eta}$.  The binning in \pt in these figures is coarser than the one used in the study. }
    \label{fig:TkTOFOffEff0}
\end{figure*}

\begin{figure*}[tbh]
 \begin{center}
  \includegraphics[width=0.90\textwidth]{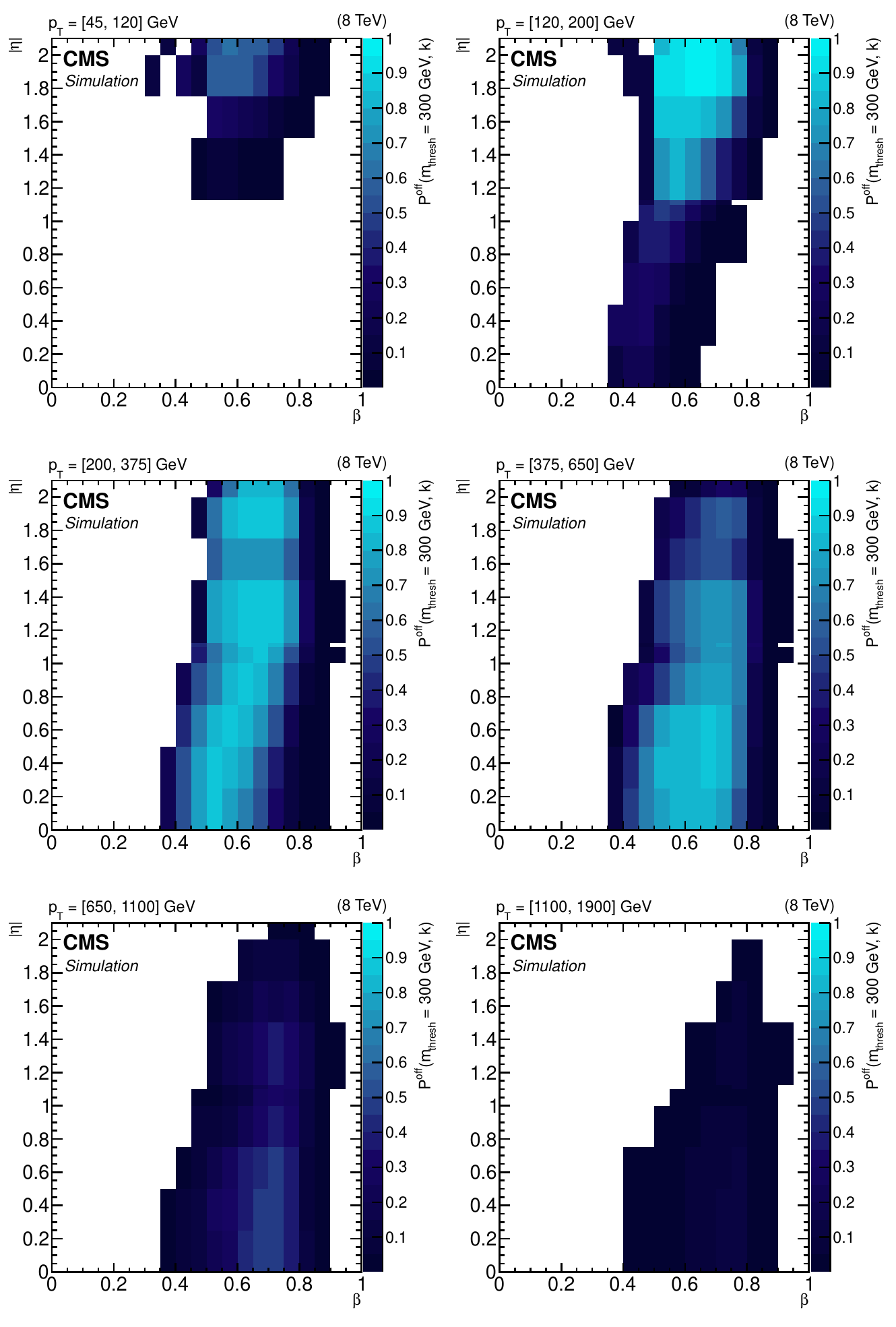}
 \end{center}
 \caption{Values taken by the probability $P^{\text{off}}(300\GeV, k)$ as a function of
the true particle-variables $\pt$, $\beta$, and $\abs{\eta}$.
 The binning in \pt in these figures is coarser than the one used in the study. }
    \label{fig:TkTOFOffEff3}
\end{figure*}

The probabilities $P^{\text{on}}(k)$ and $P^{\text{off}}(\mthresh, k)$ were evaluated in 3D bins of $\abs{\eta}$, $\beta$, and \pt.
For $\abs{\eta}$ we consider the following bin boundaries: 0.0, 0.25, 0.50, 0.75, 1.00, 1.10, 1.125, 1.50, 1.75, 2.00, 2.10.
The granularity is finer around the transition between the tracker barrel and the tracker endcap to better model the efficiency in this region.
Since the detector is symmetric in $\eta$, the probabilities for the positive and negative $\eta$ regions were averaged in order to reduce the statistical uncertainties.
A constant bin width of 0.05 is considered between 0.0 and 1.0 for the binning in $\beta$.
For the binning along \pt the following bin boundaries are considered 5, 50, 60, 70, 80, 90, 100, 110, 120, 130, 140, 150, 160, 170, 180, 190, 200, 220, 240, 260, 280, 300, 325, 350, 375, 400, 425, 450, 475, 500, 550, 600, 650, 700, 750, 800, 850, 900, 950, 1000, 1100, 1200, 1300, 1400, 1500, 1600, 1700, 1800, 1900, 2000, 2250, 2500, $\infty$\GeV.
Figures~\ref{fig:TkTOFTrigEff}, \ref{fig:TkTOFOffEff0}, and
\ref{fig:TkTOFOffEff3}  show the estimated values of $P^{\text{on}}(k)$, $P^{\text{off}}(0 \rm{\GeV}, k)$, and $P^{\text{off}}(300 \rm{\GeV},
k)$, respectively.
The complete evaluation of $P^{\text{on}}(k)$ and $P^{\text{off}}(\mthresh, k)$ for the bin ranges previously described is available as supplemental material to this paper~\cite{ProbMaps}.

The fast technique is validated by comparing the estimated signal
acceptance values with those obtained from the full
simulation for three benchmark models predicting long-lived leptons: pair
production of staus, inclusive production of staus in the context
of a gauge mediated symmetry breaking (GMSB) model, and pair production of long-lived leptons with a
null weak isospin~\cite{Langacker:2011db}.   These three models, having significantly different
kinematic properties, are studied in Ref.~\cite{EXO-12-026}.

\begin{figure*}[tbh]
 \begin{center}
  \includegraphics[width=0.49\textwidth]{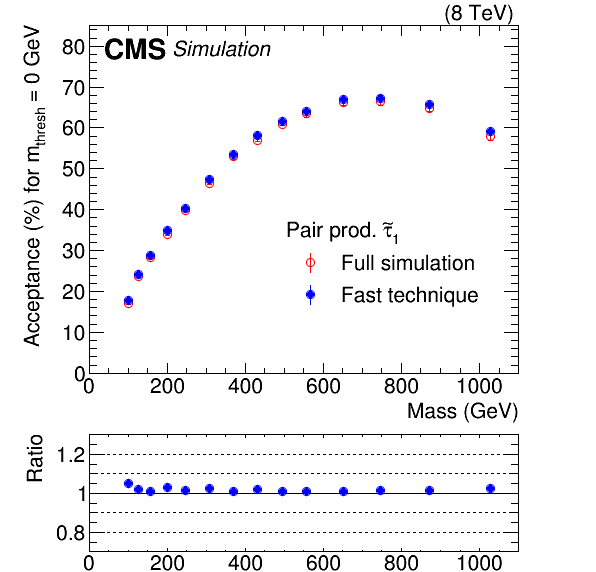}
  \includegraphics[width=0.49\textwidth]{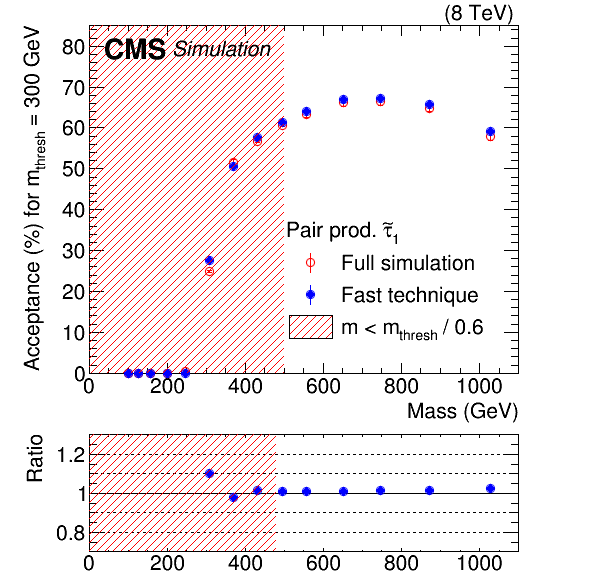} \\
  \includegraphics[width=0.49\textwidth]{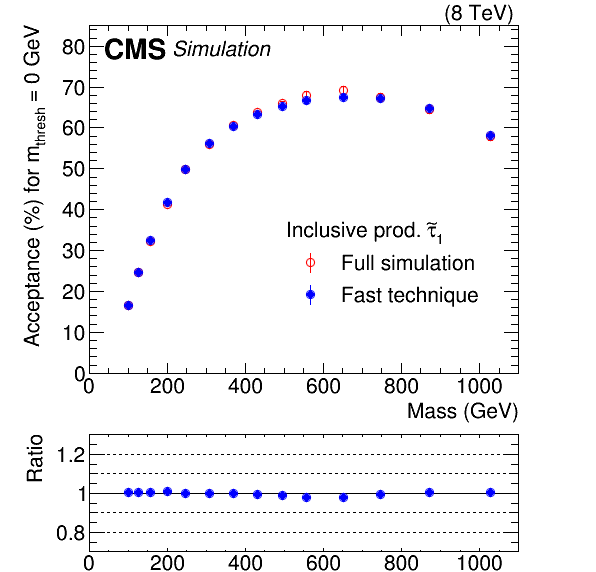}
  \includegraphics[width=0.49\textwidth]{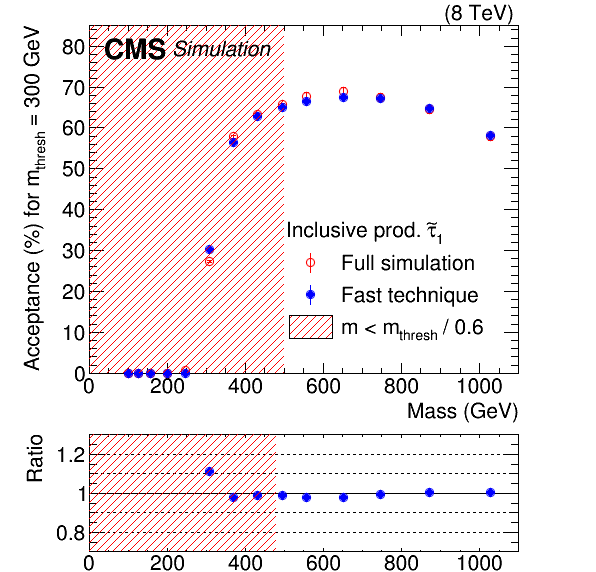} \\
 \end{center}
 \caption{Signal acceptance for a mass threshold of 0 (left) and 300\GeV (right). The upper and lower sets of distributions show the pair production and inclusive production of staus as predicted by the GMSB model, respectively.
    The panel below each figure shows the ratio of acceptance from the fast technique to the acceptance obtained from a full simulation of the detector.
    The hatched area indicates the range not satisfying the requirement on the reconstructed mass threshold.
    \label{fig:TkTOFValidationPlot1}}
\end{figure*}

\begin{figure*}[tbh]
 \begin{center}
  \includegraphics[width=0.49\textwidth]{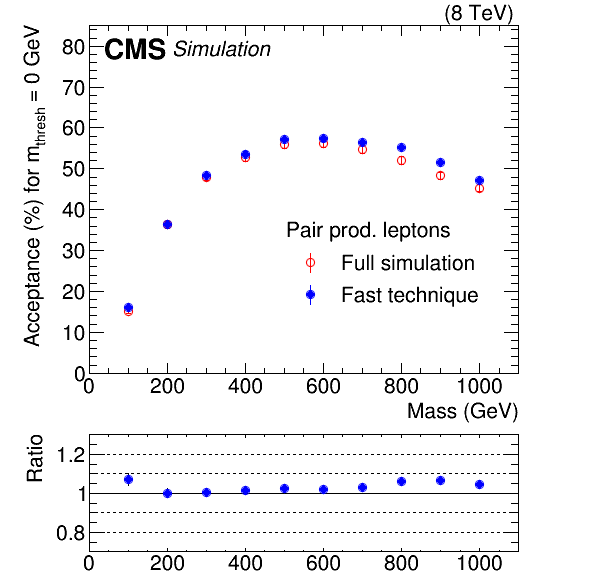}
  \includegraphics[width=0.49\textwidth]{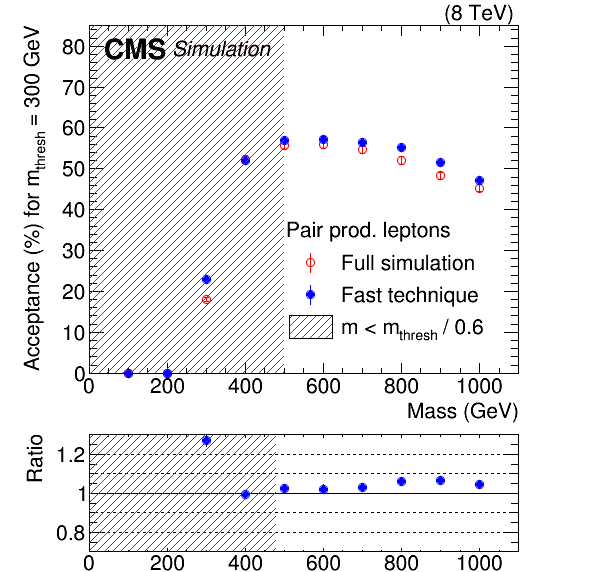} \\
 \end{center}
 \caption{Signal acceptance for a mass threshold of 0 (left) and 300\GeV (right).  The tested model is the pair production of leptons with no weak isospin.
    The panel below each figure shows the ratio of acceptance from the fast technique to the acceptance obtained from a full simulation of the detector.
    The hatched area indicates the range not satisfying the requirement on the reconstructed mass threshold.
    \label{fig:TkTOFValidationPlot2}}
\end{figure*}

The signal acceptance obtained with the two methods is shown in Figs.
\ref{fig:TkTOFValidationPlot1} and \ref{fig:TkTOFValidationPlot2} for the benchmark models considered
for two thresholds on the reconstructed mass, \mthresh.
The acceptance computed with the full simulation and with the technique presented in this paper agrees within 10\%.
The agreement between the two techniques is worse when the mass threshold is close to the true mass of the long-lived particle.
The larger disagreement close to the mass threshold is a consequence of the coarse binning of probability $P^{\text{off}}(\mthresh, k)$,
which only partially reflects the sharp variation in probability with the reconstructed values of $\pt$, $\eta$, and mass.
The reconstructed particle mass is generally lower than the actual particle mass because of a cutoff applied in the data acquisition electronics of the tracker detector.
In Ref.~\cite{EXO-12-026}, the threshold on the reconstructed mass is optimized for each signal model considered.
In this paper, for the sake of simplicity, the threshold on the reconstructed mass used in the  fast technique is set at $60\%$ of the true particle mass.
The hatched area in Figs. \ref{fig:TkTOFValidationPlot1} and \ref{fig:TkTOFValidationPlot2} indicates the range not satisfying the requirement on the reconstructed mass threshold.  Outside of this region the estimated signal acceptance is always compatible within 10\% with the results from full simulation.

\begin{figure*}
 \begin{center}
  \includegraphics[width=0.49\textwidth]{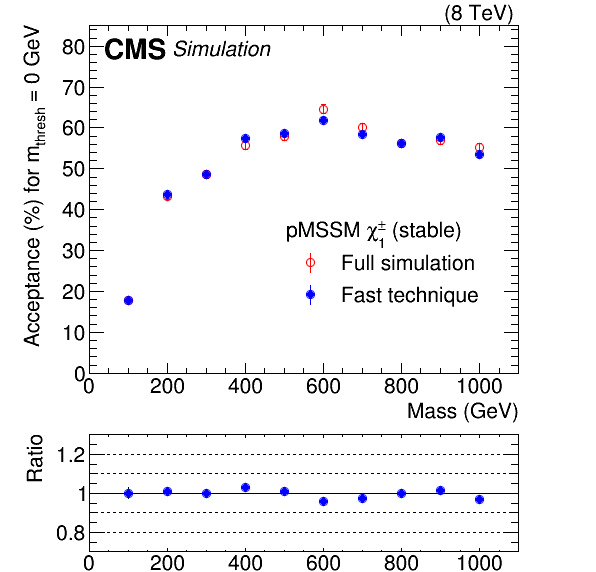}
  \includegraphics[width=0.49\textwidth]{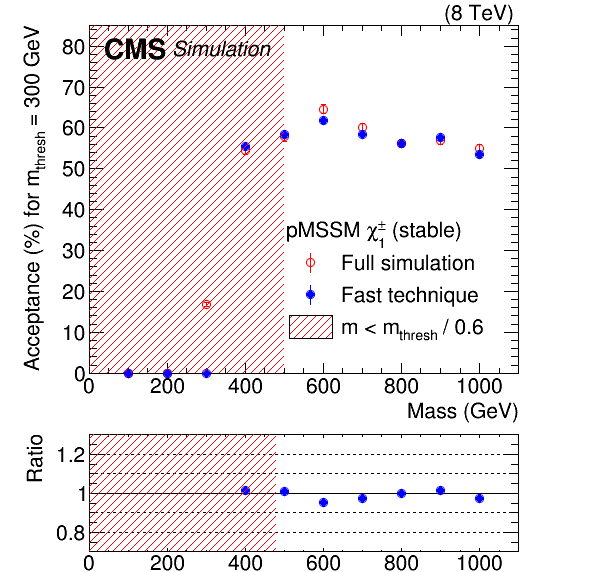}
 \end{center}
 \caption{Signal acceptance for a mass threshold of 0 (left) and 300\GeV (right) on a few representative pMSSM points predicting quasi-stable ($c\tau \geq 100$\unit{m}) charginos.
    The panel below each figure shows the ratio of acceptance from the fast technique to the acceptance obtained from a full simulation of the detector.
    The hatched area indicates the range not satisfying the requirement on the reconstructed mass threshold.
    \label{fig:TkTOFpMSSMValidation}}
\end{figure*}

Several points of the pMSSM subset have also been fully simulated, as described in Ref.~\cite{EXO-12-026},
in order to further validate the fast estimation of the acceptance
in the context of pMSSM.
The acceptances estimated using the full simulation prediction and the
proposed technique are compared in Fig.~\ref{fig:TkTOFpMSSMValidation} for points leading to quasi-stable ($c\tau \geq $100\unit{m}) charginos.  An agreement at the level of 10\% is observed.

\cleardoublepage \section{The CMS Collaboration \label{app:collab}}\begin{sloppypar}\hyphenpenalty=5000\widowpenalty=500\clubpenalty=5000\textbf{Yerevan Physics Institute,  Yerevan,  Armenia}\\*[0pt]
V.~Khachatryan, A.M.~Sirunyan, A.~Tumasyan
\vskip\cmsinstskip
\textbf{Institut f\"{u}r Hochenergiephysik der OeAW,  Wien,  Austria}\\*[0pt]
W.~Adam, T.~Bergauer, M.~Dragicevic, J.~Er\"{o}, M.~Friedl, R.~Fr\"{u}hwirth\cmsAuthorMark{1}, V.M.~Ghete, C.~Hartl, N.~H\"{o}rmann, J.~Hrubec, M.~Jeitler\cmsAuthorMark{1}, W.~Kiesenhofer, V.~Kn\"{u}nz, M.~Krammer\cmsAuthorMark{1}, I.~Kr\"{a}tschmer, D.~Liko, I.~Mikulec, D.~Rabady\cmsAuthorMark{2}, B.~Rahbaran, H.~Rohringer, R.~Sch\"{o}fbeck, J.~Strauss, W.~Treberer-Treberspurg, W.~Waltenberger, C.-E.~Wulz\cmsAuthorMark{1}
% ...
\vskip\cmsinstskip
% ...
1:~~Also at Vienna University of Technology, Vienna, Austria\\
2:~~Also at CERN, European Organization for Nuclear Research, Geneva, Switzerland\\
% ...

\end{sloppypar}
\end{document}